\documentclass[epsfig,useAMS,usegraphicx,usenatbib]{mn2e}
\usepackage{amssymb}
\usepackage{lscape} 
\title[INTEGRAL/IBIS and Swift/XRT observations of hard
cataclysmic variables]
{\emph{INTEGRAL}/IBIS and \emph{Swift}/XRT observations of hard cataclysmic variables}
\author[R. Landi et al.]
{R.~Landi$^{1}$\thanks{E-mail:landi@iasfbo.inaf.it}, L.~Bassani$^1$, A.J.~Dean$^2$, A.J.~Bird$^2$,
M.~Fiocchi$^3$, A.~Bazzano$^3$, \newauthor J.A.~Nousek$^4$, and J.P.~Osborne$^5$ \\
$^1$ INAF -- IASF Bologna, Via P. Gobetti 101, I--40129 Bologna, Italy\\
$^2$ School of Physics and Astronomy, University of Southampton,
        SO17 1BJ, Southampton, UK\\
$^3$ INAF -- IASF Rome, Via Fosso del Cavaliere 100, I--00133 Roma, Italy\\
$^4$ Department of Astronomy and Astrophysics, Pennsylvania State University, University Park, PA 16802\\
$^5$ University of Leicester, University Road, Leicester, LE1 7RH, UK}

\begin{document}

\date{}


\maketitle

\label{firstpage}

\begin{abstract}

The analysis of the third \emph{INTEGRAL}/IBIS survey has revealed several new cataclysmic variables, 
most of which turned out to be intermediate polars, thus confirming that these objects are strong 
emitters in hard X-rays. Here we present high energy spectra of all 22 cataclysmic variables detected in 
the 3rd IBIS survey and provide the first average spectrum over the 20--100 keV band for this class. Our 
analysis indicates that the best-fit model is a thermal bremsstrahlung with an average temperature of 
$<kT>$ $\sim$22 keV. Recently, eleven (ten intermediate polars and one polar) of these systems have been 
followed-up by \emph{Swift}/XRT (operating in the 0.3--10 keV energy band), thus allowing us to 
investigate their spectral behaviour over the range $\sim$0.3--100 keV. Thanks to this wide energy 
coverage, it was possible for these sources to simultaneously measure the soft and hard components and 
estimate their temperatures. The soft emission, thought to originate in the irradiated poles of 
the white dwarf atmosphere, is well described by a blackbody model with temperatures in the range 
$\sim$60--120 eV. The hard emission, which is supposed to be originated from optically thin plasma 
in the post-shock region above the magnetic poles, is indeed well modelled with a bremsstrahlung model 
with temperatures in the range $\sim$16--33 keV, similar to the values obtained from the \emph{INTEGRAL} 
data alone. In several cases we also find the presence of a complex absorber: one totally (with $N_{\rm 
H} \sim (0.4-28) \times 10^{21}$ cm$^{-2}$) and one partially (with $N_{\rm H} \sim (0.7-9) \times 
10^{23}$ cm$^{-2}$) covering the source. Only in four cases (V709 Cas, GK Per, IGR J06253+7334 and IGR 
J17303--0601), we find evidence for the presence of an iron line at 6.4 keV. We discuss our findings in 
the light of the systems parameters and cataclysmic variables/intermediate polars modelling scenario.

\end{abstract}


\section{Introduction}

Cataclysmic variables (CVs) are close binary systems, with orbital period typically less than one day, 
containing a white dwarf (WD) which is accreting material from a late-type main sequency secondary star 
filling its Roche lobe (for a review see Warner 1995). As the accreting matter falls onto the WD, it 
forms a disk; the infalling matter releases its gravitational energy, heating the accretion disk and the 
regions where matter falls onto the WD. As a result, radiation is emitted over many frequencies, from 
infrared to X-rays. Recently, this class of binaries has re-gained interest because of their detection 
in large numbers at energies above 20 keV, where very few cases were previously known thanks to 
\emph{RXTE} (Suleimanov et al. 2005) and \emph{Beppo}SAX (De Martino et al. 2004b) observations. CVs can 
be broadly divided into two subclasses: non magnetic and magnetic objects depending on the strength of 
the WD magnetic field.

In non-magnetic CVs (e.g. Dwarf Novae or DNe) the hard X-ray emission originates from the boundary layer 
between the accretion disk and the WD surface and depends on the mass accretion rate ($\dot{M}$) of the 
disk (Pringle \& Savonije 1979). This boundary can release up to $50\%$ of the total accreting energy 
when the material decelerates on to the WD surface. For low $\dot{M}$ ($\lesssim 10^{16}$ g 
s$^{-1}$) the boundary is optically thin and emits hard X-rays with temperatures up to $10^{8}$ K. 
For high accretion rate ($\gtrsim 10^{16}$ g s$^{-1}$) the boundary layer is thick, thus suppressing 
hard X-rays and emitting mostly in soft X-rays with temperature $\sim$$10^{5}$ K (Pringle \& Savonije 
1979).

Magnetic cataclysmic variables (mCVs) are divided in two classes: polars (or AM Her type) and 
intermediate polars (IPs or DQ Her type). In polars (for a review see Cropper 1990) the magnetic field 
is so strong ($B > 10^{7}$ Gauss, De Martino et al. 2004a) that it forces the WD to spin with 
the same period of the binary system ($P_{\rm rot,WD}=P_{\rm orb}$) and preventing the formation of an 
accretion disk around the WD. The infalling material is chanelled by the magnetic field along its lines 
and falls on the magnetic poles of the WD. In the IPs, instead, the weaker magnetic field ($B \sim 
10^{6}-10^{7}$ Gauss, De Martino et al. 2004a) does not synchronize the spin period of the WD with 
the orbital period of the binary system ($P_{\rm rot,WD} < P_{\rm orb}$). In these objects, the 
accretion process is through a disk , which is truncated in its inner region because of the interaction 
with the WD magnetosphere; the overall result is the formation of an accretion curtain rather than a 
converging stream. In these systems, it is also likely that part of the material from the companion can 
flow directly on the magnetic poles without passing through the disk (disk-overflow, Hellier 1991).

In magnetic CVs the broad-band X-ray spectrum is mainly due to the superposition of a soft and a hard 
component. The hard X-ray emission is the consequence of the shock produced in the accreting material as 
it impacts the WD atmosphere. The hot post-shock region cools via thermal bremsstrahlung process 
emitting hard X-rays; the temperatures of the post-shock region are typically in the range $\sim$5--60 
keV (Warner 1995; Hellier 2001). This emission is highly absorbed ($N_{\rm H} \sim10^{23}$ 
cm$^{-2}$) within the accretion flow (Ishida et al. 1994) and is expected to be reflected by the WD 
surface (De Martino et al. 2004b and references therein). In polars, due to the strong magnetic field, 
cyclotron radiation cooling suppresses the high temperature bremsstrahlung emission of a substantial 
fraction of the electrons in the shock region (Lamb \& Masters 1979; K\"{o}nig, Beuermann \& 
G\"{a}nsicke 2006). This may explain why most of hard X-ray detected CVs are IPs, in which cyclotron 
emission is negligible.

The soft X-ray component, originating by the reprocessing of the hard emission on the WD surface, is 
typically described by a blackbody emission with temperatures ranging from a few eV up to $\sim$100 eV 
(De Martino et al. 2004, 2006a,b). This component was mostly seen in polars but recently Evans 
\& 
Hellier (2007), performing a systematic spectral analysis of \emph{XMM-Newton} data of IPs, found this 
to be a common feature of IPs X-ray spectra. Furthermore, these authors suggest that the lack of a 
blackbody component in some of these systems is to be ascribed to geometrical effects: the accretion 
polar regions are either widely hidden by the accretion curtain or are only visible when foreshortened 
on the WD limb (depending on the inclination of the system).

Thanks to \emph{INTEGRAL} observations strategy, a large number of new CVs have now been detected and 
classified through optical spectroscopy (Masetti et al. 2006,2008; Bikmaev et al. 2006; G\"{a}nsicke et 
al. 2005). In the third \emph{INTEGRAL}/IBIS survey, 22 CVs are reported, confirming this class of 
binaries to be strong emitters in hard X-rays. In this paper we present IBIS spectral analysis of the 
entire sample of CVs detected by \emph{INTEGRAL}. For eleven systems, which have been observed by 
\emph{Swift}/XRT, we also report the broad-band ($\sim$0.3--100 keV) analysis by combining not 
simultaneous XRT and IBIS data. A detailed investigation of the temporal properties and of the spectral 
behaviour of the pulsations (i.e. the spectral analysis at the maximum and minimum of the pulse cycle) 
is beyond the aim of this paper. Therefore, here we only present the spectral analysis of the 
phase-averaged spectra.

\section{Observation and data analysis}

\subsection{\emph{INTEGRAL}}

The \emph{INTEGRAL} data reported here consist of several pointings performed by the IBIS/ISGRI 
instrument (Ubertini et al. 2003) between revolution 12 and 429, i.e. the period from launch to the end 
of April 2006. \emph{ISGRI} images for each available pointing were generated in various energy bands 
using the ISDC offline scientific analysis software OSA v. 5.1 (Goldwurm et al. 2003). Count rates at 
the position of the source were extracted from individual images in order to provide light curves in 
various energy bands; from these light curves, average fluxes were then obtained and combined to produce 
an average source spectrum (see Bird et al. 2007 for details). Spectra were produced in nine energy bins 
over the 20--100 keV band. An appropriately re-binned \textit{rmf} file was also produced from the 
standard IBIS spectral response file to match the \textit{pha} file energy bins. Here and in the 
following, spectral analysis was performed with XSPEC v. 11.3.2 package (Arnaud 1996) and errors are 
quoted at 90\% confidence level for one interesting parameter ($\Delta\chi^{2}=2.71$).

In Table~\ref{Tab1} we list all 22 CVs detected so far by \emph{INTEGRAL} together with their 
position and classification. As can be seen, this sample contains mainly magnetic systems (21 objects), 
with the majority ($\sim$81\%, or 17 versus 4) being IPs (10 systems confirmed and seven 
possible/probable IP candidate, as reported in Table~\ref{Tab1}). Only one system (SS Cyg) is classified 
as DNe.

\begin{figure}
\includegraphics[width=1.0\linewidth]{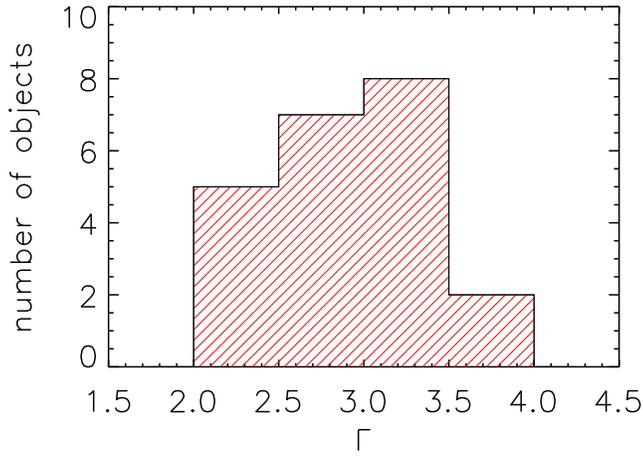}
\caption{Photon index distribution of the whole IBIS CVs sample.}
\label{fig1}
\end{figure}

\begin{figure}
\includegraphics[width=1.0\linewidth]{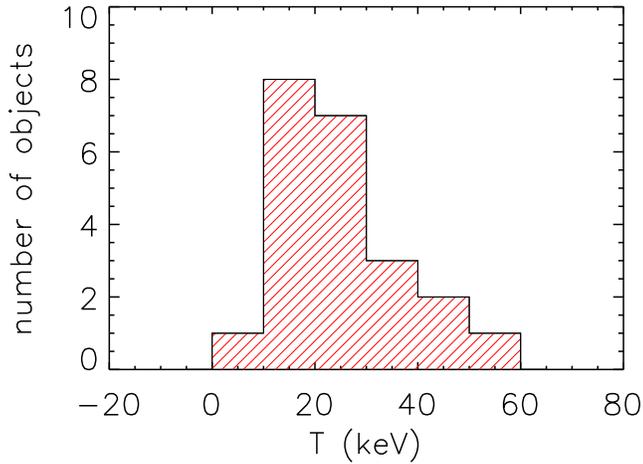}
\caption{Temperature distribution of the whole IBIS CVs sample.}
\label{fig2}
\end{figure}

\begin{figure}
\includegraphics[width=0.70\linewidth,angle=-90]{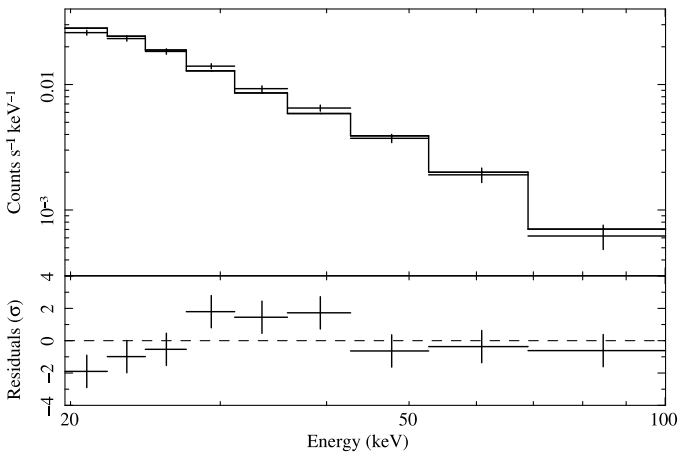}
\caption{IBIS average spectrum fitted with a simple power law   
(upper panel); residuals to this model are in units of $\sigma$ (lower panel).}
\label{fig3}
\end{figure}

\begin{figure}
\includegraphics[width=0.70\linewidth,angle=-90]{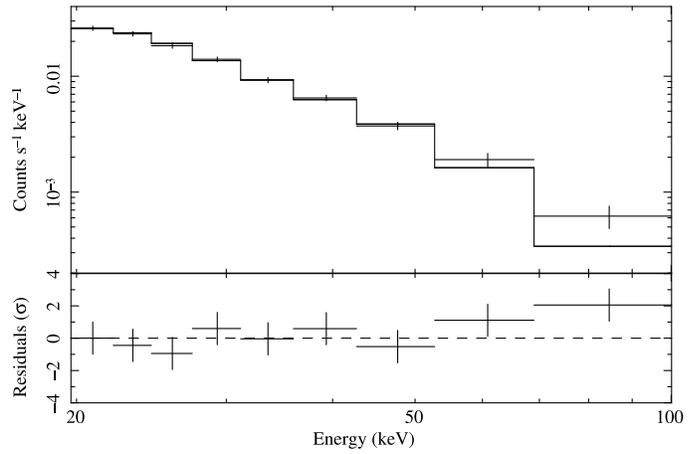}
\caption{IBIS average spectrum fitted with a bremsstrahlung model
(upper panel); residuals to this model are in units of $\sigma$ (lower panel).}
\label{fig4}
\end{figure}

To characterise the hard X-ray emission (20--100 keV) of these systems, we fit the IBIS spectra with two 
models, a simple power law and a thermal bremsstrahlung. The results of these fits are reported in 
Table~\ref{Tab1}, where $\Gamma$, $kT$, 20--100 keV flux (assuming a bremsstrahlung model) and 
$\chi^{2}/\nu$ relative to both models are listed for each source. Generally, both models provide an 
acceptable fit, although in a few cases the bremmsstrahulung gives a better $\chi^{2}/\nu$. Inspection 
of Table~\ref{Tab1} also indicates that there are no significant differences between the spectral 
parameters of the different classes of CVs. The photon index and temperatures distributions (shown in 
Figure~\ref{fig1} and Figure~\ref{fig2}) peak at $\Gamma = 2.9\pm0.5$ and $kT = 26.0\pm11.0$ keV, 
respectively, in substantial agreement with what reported previously by Barlow et al. (2006) for a 
subsample. Thus, in order to improve the statistics, we fitted together all data sets allowing the flux 
to vary from source to source, but imposing the slope and the bremsstrahlung temperature to be the same 
for all objects. These fits yield a temperature of $kT = 22.0\pm0.9$ keV and a photon index $\Gamma = 
2.96\pm0.05$, much in agreement with the mean values estimated from Figure~\ref{fig1} and 
Figure~\ref{fig2}. Comparison of the $\chi^{2}/\nu$ strongly suggests that the thermal bremsstrahlung is 
the model that better reproduces the data ($\chi^{2}/\nu=272.6/175$) with respect the power law 
($\chi^{2}/\nu=307/175$); this is reflected in the residuals to the models which are shown in 
Figure~\ref{fig3} and Figure~\ref{fig4}. This result supports the hypothesis that the hard X-ray 
emission is due to a thermal bremsstrahlung component.

\subsection{\emph{Swift}/XRT}

For eleven sources in our sample, we also have X-ray observations acquired with the XRT (X-ray 
Telescope, 0.2--10 keV, Burrows et al. 2005) on board the \emph{Swift} satellite (Gehrels et al. 2004). 
XRT data reduction was performed using the XRTDAS standard data pipeline package ({\sc xrtpipeline} v. 
0.10.6), except for the observations performed after August 2007 (in these cases we used the latest 
version 0.11.6 of the pipeline), in order to produce screened event files. All data were extracted only 
in the Photon Counting (PC) mode (Hill et al. 2004), adopting the standard grade filtering (0--12 for 
PC) according to the XRT nomenclature. The log of all X-ray observations, available at June 2 2008, is 
given in Table~\ref{Tab2}. For each measurement, we report the XRT observation date, the total on-source 
exposure time and the total net count rate in the 0.3--10 keV energy band.

For those sources in which the XRT count rate was high enough to produce event pile-up, we extracted the 
events in an annulus centered on the source, by determining the size of the inner circle according to 
the procedure described in Romano et al. (2006). In the other cases, events for spectral analysis were 
extracted within a circular region of radius 20$^{\prime \prime}$, centered on the source position, 
which encloses about 90\% of the PSF at 1.5 keV (see Moretti et al. 2004).

The background was taken from various source-free regions close to the X-ray source of interest, using 
circular regions with different radii in order to ensure an evenly sampled background. In all cases, the 
spectra were extracted from the corresponding event files using the {\sc XSELECT} software and binned 
using {\sc grppha} in an appropriate way, so that the $\chi^{2}$ statistic could be applied. We used 
version v.009 of the response matrices and create individual ancillary response files \textit{arf} using 
{\sc xrtmkarf v. 0.5.6}.

We summed together all the available observations for each source to provide an average source spectrum 
as done for the \emph{INTEGRAL}/IBIS data. A posteriori, this approach proved resonable since only in 
two cases (V709 Cas and GK Per) we observed significant changes in flux, by a factor of 1.5 and 1.7 
respectively, but not in shape.

\section{The broad-band spectral analysis}

Use of broad-band data is particularly important to determine the properties of the emitting regions, in 
particular the temperature of the optically thin plasma, the presence of interstellar and/or 
circumstellar absorption; it is also useful to study the effects of reprocessing of the hard X-ray 
emission on the WD surface. The basic model we adopted consists of a blackbody ($kT_{\rm BB}$) and a 
bremsstrahlung ($kT_{\rm Brems}$) component to account for the soft and hard X-ray emissions expected in 
magnetic CVs, plus two cold media consisting of a simple absorber, totally covering the source (with a 
column density $N_{\rm H}$) and a partial covering absorber (with a column density $N_{\rm H(pc)}$ and 
covering factor $C_{\rm F}$). If required by the data, we also included in the model another emission 
component consisting of an optically thin plasma (MEKAL code in XSPEC) and a K$\alpha$ iron line. We 
also tested the data for the presence of a multi-temperature structure of the post-shock region and of a 
reflection component, but the data were of too low statistical quality to allow these more complex fits. 
We introduced in the fitting procedure a cross-calibration constant ($C_{\rm calib}$) to account for 
possible mismatch between XRT and IBIS data as well as for source flux variations. The results of the 
spectral fits including cross-calibration constant and reduced $\chi^{2}$ are all listed in 
Table~\ref{Tab3}; spectra and residuals with respect to the best-fit model are plotted in 
Figure~\ref{fig5}.
We found that in most cases our basic model provides a good description of the broad-band spectra of 
our CVs. The cross-calibration constants have been found to be consistent, within the uncertainties, 
with unity for all sources apart from GK Per, in which the low value of the constant can be ascribed to 
the flux variability of the source, as confirmed by XRT (this work) and previous observations 
(Vrielmann, Ness \& Schmitt 2005 and references therein). Three systems out of eleven (XSS J12270-4859, 
IGR J16500--3307 and IGR J17195--4100) do not show evidence for complex absorption, but this could be 
due to the short exposure of the XRT observations. In eight cases (V709 Cas, IGR J06253+7334, XSS 
J12270--4859, IGR J14536--5522, IGR J15479--4529, IGR J16167--4957, IGR J17195--4100 and IGR 
J17303--0601) the column density of the fully covering absorber is lower or compatible with the Galactic 
absorption in the direction of the sources, suggesting an interstellar origin for this component. For 
systems showing the partial covering absorption, we found that the absorber is remarkably higher than 
the galactic one, indicating an intrinsic origin likely associated to the presence of the accretion 
material along the line of sight. The range of values found for both total and partial absorber are 
consistent with what found by Evans $\&$ Hellier (2007) by analysing the \emph{XMM-Newton} data of 
several IPs\footnote{The reader should note that the partial covering model is likely due to time 
averaging of a variable absorption as the WD and/or orbit rotate.}. Only in four cases (V709 Cas, GK 
Per, IGR J06253+7334 and IGR J17303-0601) we find evidence for the presence of an iron line at around 
6.4 keV; lack of iron line detection in the other sources may be due to the lower statistical quality of 
their X-ray data.

An interesting result of our analysis is that most of the IPs of our sample significantly require a 
blackbody component (see Table~\ref{Tab4}); the temperatures related to this component are all in the 
range 60--120 eV as observed in previous studies (Evans \& Hellier (2007); De Martino et al. 
2001,2004b,2006b,2008). Following Evans $\&$ Hellier (2007), we would not expect to observe a blackbody 
component if the partial covering absorption is associated to the accretion curtain; however, in our 
case the blackbody emission does not seem to be affected by the strength of the partial absorber and is 
observed in all CVs. The fact that in FO Aqr we can determine the blackbody temparature, while Evans \& 
Hellier (2007) do not, may be related to our broad-band analysis which allows a better definition of the 
source continuum and an estimate of the soft component temperature. A bremmsstrahulung component is 
required in every object and its temperature ranging from 16--35 keV, is generally consistent with 
values obtained using IBIS data alone.

\section{Notes on a few individual sources} 
{\bf V709 Cas}: in this case our results are in agreement 
with what found by Falanga et al. (2005) from a joint JEM-X/ISGRI spectrum. We confirm a temperature for 
the hard X-ray emission at around 25 keV, without claiming the presence of a reflection component as 
instead reported by De Martino et al. (2001), even if we fit the data assuming their model. In this case 
the data shows an excess at around 6.4 keV, that we modeled by adding an iron line to the basic model 
(see Table~\ref{Tab3}). The addition of this component is required by the data at 99.99$\%$ confidence 
level and provides an energy centroid $E =6.46^{+0.05}_{-0.04}$ keV and an $EW = 140^{+46}_{-55}$ eV, 
which is consistent with the value reported by De Martino et al. (2001).\\
{\bf GK per}: a recent analysis of \emph{XMM-Newton} data of GK Per has revealed the presence of 
spectral complexity in the soft energy band, which has been resolved in a multitude of emission lines 
thanks to the high-resolution spectroscopy performed with the RGS (Vrielmann et al. 2005). Furthermore, 
the image of the source taken with the ACIS-S detector on board \emph{Chandra}, reported by the same 
authors, shows the presence of extended emission (a shell with a radius of $\sim$$50$ arcsec) around 
the source, which disappears above $\sim$1.5 keV. With the XRT 
data we cannot evaluate the contribution of this component to the total emission, therefore we decided 
to perform the spectral analysis above $\sim$1 keV. The fit with the basic model is not satisfactory 
($\chi^{2}/\nu = 693.2/502$): the residuals to the model show a clear excess of counts around $\sim$6 
keV. The iron K$\alpha$ region of this source has been recently resolved by \emph{Chandra} in three 
distinct components (6.4, 6.7 and 6.97 keV, Hellier \& Mukai (2004)), with the fluorescent line showing 
a red wing up to 6.33 keV. This last component has been confirmed by \emph{XMM-Newton} observations 
(Vrielmann et al. 2005). In our analysis, the addition of a Gaussian Fe K$\alpha$ line is strongly 
required by the data ($> 99.99\%$, $\Delta\chi^{2}= 183$ for three degrees of freedom) and gives an 
energy centroid $E = 6.35^{+0.05}_{-0.03}$ keV, a line width $\sigma = 0.29^{+0.05}_{-0.04}$ keV and an 
equivalent width $EW = 371^{+59}_{-53}$ eV, in agreement with what reported by Vrielmann et al. (2005). 
The confidence contours of the line energy versus line width are shown in Figure~\ref{fig6}. 
Unfortunately, because of the low sensitivity of XRT in this energy band, we cannot further resolve this 
spectral region to verify the presence of other lines and/or of a red wing in the neutral iron line.\\ 
{\bf IGR J06253+7334}: in this case, although the basic model provides a good description of the data 
($\chi^{2}/\nu = 111.8/121$), an excess at around 6.4 keV is clearly evident. The addition of a narrow 
Gaussian line at 6.4 keV is required by the data at 99.5\% confidence level ($\Delta\chi^{2}= 7.3$ for 
one degree of freedom) and provides an $EW = 195^{-108}_{+115}$ eV (see Table~\ref{Tab3}). By leaving 
the iron line energy free to vary, we found an energy centroid $E=6.41^{+0.06}_{-0.04}$, a similar $EW$, 
but a lower significance of the line (98.4\% confidence level). Concerning the temperature of the 
bremsstrahlung component, we found a difference between the value found by fitting the IBIS data alone 
($\sim$7 keV) and that found in the broad-band spectral analysis ($\sim$35 keV). This discrepancy could 
be due to a source variation between XRT and IBIS observations, or perhaps it was the use of a more 
complex model which resulted in the different parameter value.\\
{\bf IGR J17303--0601}: also for this source the basic model does not reproduce satisfactorily the 
spectrum ($\chi^{2}/\nu = 358/292$). The addition of a further emission component consisting of an 
optically thin plasma (MEKAL code in XSPEC) improves the fit ($\Delta\chi^{2}= 38$ for two degrees of 
freedom) and is significant at 99.99\%, giving a temperature $kT = 0.20^{+0.03}_{-0.02}$ keV. The 
addition of a narrow Gaussian line at 6.4 keV to account for the excess at around 6 keV is only 
marginally required by the data ($\sim$$95\%$) and provides an $EW = 83^{+60}_{-65}$ eV. Our results are 
consistent with those recently reported by De Martino et al. (2008) from a combined analysis of 
\emph{XMM-Newton} and \emph{INTEGRAL} data, except for the temperature of the hard X-ray emission (which 
is lower in our fit than in theirs) and the metal abundances, which we freeze to the solar value because 
of the lack of statistics.

\section{Discussion} 

We have analysed \emph{INTEGRAL} data of a sample of 22 hard X-ray selected CVs, the majority of 
which are IPs. We confirm previous indications that the high energy spectra of these type of objects 
are well described by either a bremsstrahlung model with $kT$ in the range 7--56 keV or a power law 
with $\Gamma$ in the range 2.1--3.5; however, the average \emph{INTEGRAL} spectrum of all 22 CVs is 
significantly better fitted by the bremsstrahlung model with an average temperature of 22 keV. 
Confirmation that IPs are the most powerful soft gamma-ray emitters within the CV population is an 
important result defined from \emph{INTEGRAL} observations. The lack of emission above a few 
tens of keV from a large set of non-magnetic white dwarf binaries and synchronised polars is also an 
interesting information. For non-magnetic CVs the non detection of high energy emission could be 
related to the
lack of radial accretion flow. Instead, the fact that IPs are high energy emitters and polars are not
suggests that
the onset of emission above 10 keV requires some level of synchronisation of 
the orbital and spin periods, but then the high energy radiation is less likely to be observed when 
full synchronisation sets in.
One would then expect to observe some level of correlation between the 
characteristics of the soft gamma-ray emission and the degree of synchronisation. In 
Table~\ref{Tab5} we 
report for each source in the complete sample of CVs those parameters which characterise the system; 
data are obtained from the literature\footnote{For semplicity, for many of our sources, we refer to the 
work of Barlow et al. 2006, which contains all the original references; similar information can also be 
found in Ritter \& Kolb (2003).} (see references in Table~\ref{Tab5}) and 
include the orbital 
and spin periods and their ratio, the WD mass and the source distance. Not all sources are fully 
characterised and in the cases of newly discovered \emph{INTEGRAL} sources data are often unavailable. 
As a first step we cross-correlated the bremsstrahlung temperature and 20--100 keV flux as obtained in 
Table~\ref{Tab1} with the various parameters listed in Table~\ref{Tab5}, but have not found any 
convincing correlations. In particular, neither of these two gamma-ray parameters cross correlate with 
the ratio of spin to orbital period: while a wide range (extending from around 0.02 to up $\sim$1) in 
the degree of synchronisation is apparent in the sample studied, yet there is a notable similarity in 
the thermal bremsstrahlung temperatures throughout the sample. It remains to understand how systems 
which are quite different in various parameters are able to privide such a well confined range of 
bremsstrahlung temperatures.

Half of the sources in the sample have also X-ray data available so that the low energy emission could 
be parametrized in terms of a blackbody model with well constrained temperatures in the range 60--120 
eV. Extra features to this baseline model have been found in a few sources such as an iron line at 6.4 
keV in four sources and an extra thermal component in one object. Both soft and hard components are 
viewed through two layers of neutral absorbing material: the first seen in all 11 objects totally covers 
the emitting region, while the second, seen in $\sim$70\% of them, only partially covers the source. We 
also searched for correlation between the system values and soft X-ray emission parameters but again 
found none. The measured X-ray (2--10 keV) and soft gamma-ray (20--100 keV) luminosities are found to 
span from 10$^{32}$ to 10$^{33}$ erg s$^{-1}$; their ratios span from 0.02 to 7.5 with an average at 
around 1.3, but only when GK Per, with a ratio of $\sim$38, is discounted. GK Per stands out in the 
sample because it has a considerably longer orbital period than the rest of the group and, by a 
considerable margin, the highest X-ray to gamma-ray flux ratio. GK Per was a bright nova occurring in 
1901 in Perseus and is now classified as an IP. It is possible that the system is still quite young and 
has not fully switched into the standard IP state as described here; coverage in time would be extremely 
interesting to see if any change towards a more standard IP behaviour will take place. The bolometric 
luminosities of the blackbody emitting components have corresponding emitting areas in the range 
$10^{12}$ to $\sim$$10^{16}$ cm$^{2}$, with GK Per and FO Aqr located at the top end of the range. From 
the available values of the WD mass, we estimate that the observed blackbody emitting area covers from 
$\sim$$3\times10^{-3}$ to $\sim$$9\times10^{-7}$ of the WD surface.

Finally, it is important to underline that a unique baseline model has been able to describe the 
broad-band (0.3--100 keV) data of half the sample of \emph{INTEGRAL} detected objects. This 
model is consistent with the standard IP scenario, where thermal emission originating in the 
irradiated poles of the WD atmosphere provides the soft X-ray component, while emission in the 
post-shock region above the magnetic poles gives rise to the thermal bremsstrahlung component; this 
emission is absorbed within the accretion flow and is possibly reflected by the WD surface, hence 
the production of iron line features. This baseline model can be used as a pointer to IP 
classification (Ramsay et al. 2008) and considered a signature of such systems within the CVs 
population.

\section*{Acknowledgments} 
Based on observations with INTEGRAL, an ESA project with instruments and 
science data centre funded by ESA member states (especially the PI countries: Denmark, France, Germany, 
Italy, Switzerland, Spain), Czech Republic and Poland, and with the participation of Russia and the USA. 
We acknowledge the following funding: in Italy, Italian Space Agency financial and programmatic support 
via contracts I/088/06/0 and I/008/07/0; in the UK, funding via PPARC grant PP/C000714/1. We acknowledge 
the use of public data from the Swift data archive.

\clearpage

\begin{table*}
\begin{center}
\footnotesize
\caption{Characteristics of the CVs belonging to the third IBIS Survey and their best-fit IBIS parameters
by assuming power law and bremsstrahlung models.}
\label{Tab1}
\begin{tabular}{lcccccccc}
\hline
\hline
Source & RA & Dec & Class$^{a,b}$ & $\Gamma$ & $\chi^{2}/\nu$ & kT$_{\rm Brems}$ & $\chi^{2}/\nu$ 
& F(20--100 keV)$^{c}$\\ 
       & (J2000)  & (J2000) &  &  &  & (keV) &  &  ($10^{-11}$ erg cm$^{-2}$ s$^{-1}$)  \\
\hline
\hline
IGR J00234+6141   &     5.726   &    +61.706   &  IP$^{\bullet}$   & $2.18^{+1.24}_{-1.04}$ & 14.0/7 &
$>$14 & 13.8/7 & 0.6 \\
V709 Cas          &     7.204   &    +59.306    &  IP$^{\bullet}$  &  $2.77^{+0.13}_{-0.12}$ & 13.2/7 &
$25.5^{+3.1}_{-2.7}$ & 3.9/7  &  5.0 \\
GK Per            &    52.778   &    +43.934   &  IP$^{\bullet}$  & $2.21^{+0.84}_{-0.72}$ & 4.6/7 & 
$43.7^{+128.8}_{-23.4}$ & 3.6/7  &  2.1 \\
BY Cam            &    85.737   &    +60.850    &  P  & $2.40^{+1.08}_{-0.87}$ & 5.4/7 & 
$36.7^{+230.3}_{-22.7}$ & 5.9/7  &  3.2 \\
IGR J06253+7334   &    96.340   &    +73.602     &  IP$^{\bullet}$  & $> 3.5$ & 2.2/7 &
$7.0^{+8.9}_{-3.8}$ & 2.2/7  &  0.7 \\
XSS J12270--4859  &   187.007   &   --48.893   &  IP$^{\bullet \bullet \bullet}$  & 
$2.29^{+0.49}_{-0.44}$ & 9.1/7 & 
$42.5^{+36.0}_{-15.8}$ & 6.6/7   &  2.5  \\
V834 Cen          &   212.249   &   --45.273   &   P  & $3.00^{+1.77}_{-1.28}$ & 6.8/7 &
$18.7^{+61.0}_{-10.6}$ & 6.8/7   &  0.7\\
IGR J14536--5522  &   223.435   &   --55.374   &   P  & $3.58^{+0.61}_{-0.52}$ & 7.1/7 & 
$14.0^{+5.4}_{-3.7}$ & 8.1/7  &  1.4 \\
IGR J15094--6649  &   227.351   &   --66.844 & IP$^{\bullet \bullet \bullet}$  & $3.16^{+0.67}_{-0.55}$ 
& 3.8/7 & $18.8^{+11.5}_{-6.5}$ & 6.5/7   &  1.7\\
IGR J15479--4529  &   237.050   &   --45.478     &  IP$^{\bullet}$  & $2.64^{+0.13}_{-0.12}$ & 29.3/7 &
$29.3^{+3.1}_{-2.7}$ & 6.2/7   &  6.1\\
IGR J16167--4957  &   244.140   &   --49.974  & IP$^{\bullet \bullet \bullet}$  & 
$3.28^{+0.27}_{-0.25}$ & 10.5/7 &   $17.3^{+3.3}_{-2.7}$ & 10.0/7   &  2.3\\
IGR J16500--3307  &   252.491   &   --33.064 & IP$^{\bullet \bullet,d}$  & $3.18^{+0.47}_{-0.40}$ & 
18.8/7 &
$20.0^{+6.9}_{-4.7}$ & 14.5/7   &  1.5\\  
V2400 Oph         &   258.172   &   --24.280    &  IP$^{\bullet}$  & $3.29^{+0.15}_{-0.14}$ & 13.7/7 &
$16.6^{+1.9}_{-1.7}$ & 8.7/7  &  3.3 \\
IGR J17195--4100  &   259.906   &   --41.014     &  IP$^{\bullet \bullet}$  & $2.86^{+0.19}_{-0.18}$ & 
8.8/7 &     
$25.0^{+4.6}_{-3.6}$ & 19.4/7  &  3.5 \\
IGR J17303--0601   &   262.596   &    --5.988    &  IP$^{\bullet}$  & $2.56^{+0.32}_{-0.29}$ & 7.5/7 &      
$32.4^{+12.5}_{-8.1}$ & 8.8/7  &  4.5 \\
V2487 Oph         &   262.963   &   --19.233  & IP$^{\bullet \bullet \bullet}$ & $2.12^{+0.51}_{-0.47}$ 
& 4.4/7 &   $55.6^{+76.4}_{-24.5}$ & 3.6/7  &  1.2\\
V1223 Sgr         &   283.755   &   --31.154     &  IP$^{\bullet}$  & $3.26^{+0.13}_{-0.12}$ & 10.5/7 &      
$17.7^{+1.6}_{-1.4}$ & 3.1/7   &  7.1\\
V1432 Aql  &   295.058   &   --10.428    &   P  & $2.82^{+0.56}_{-0.49}$ & 4.8/7 &      
$24.4^{+15.3}_{-8.1}$ & 5.1/7  &  3.4 \\
V2069 Cyg         &   320.906   &    +42.278  & IP$^{\bullet \bullet}$  & $3.20^{+0.76}_{-0.60}$ & 
4.6/7 &  $19.2^{+12.3}_{-6.8}$ & 4.5/7  &  1.2\\
IGR J21335+5105   &   323.438   &    +51.121     &  IP$^{\bullet}$  & $2.91^{+0.25}_{-0.24}$ & 22.0/7 &
$23.6^{+5.0}_{-4.0}$ & 12.4/7   &  4.1\\
SS Cyg            &   325.691   &    +43.583   &  DN  & $3.19^{+0.27}_{-0.25}$ & 2.7/7 &  
$18.4^{+3.7}_{-2.9}$ & 4.8/7 &  3.7\\ 
FO Aqr            &   334.478   &    --8.317    &  IP$^{\bullet}$  & $2.70^{+1.28}_{-0.89}$ & 7.5/7 & 
$29.7^{+70.1}_{-16.6}$ & 7.8/7 &  4.2 \\   
\hline
\hline 
\end{tabular}
\begin{list}{}{}
$^{a}$ IP: intermediate polar; P: polar; DN: dwarf nova;\\
$^{b}$ The classification of the IP systems is addressed following the Koji Mukai's IP probability 
classification (available at http://asd.gsfc.nasa.gov/Koji.Mukai/iphome/catalog/alpha.html): 
$^{\bullet}$:
confirmed; $^{\bullet \bullet}$: probable; $^{\bullet \bullet \bullet}$: possible;\\ 
$^{c}$ The flux is derived by assuming the bremsstrahlung model;\\
$^{d}$ For this source we used the classification proposed by Masetti et al. (2008). 
\end{list}
\end{center}
\end{table*}

\begin{table*}
\begin{minipage}{155mm}
\centering
\caption{Log of the \emph{Swift} XRT observations used in this paper (until June 2, 2008).}
\label{Tab2}
\begin{tabular}{lccc}
\hline
\hline
Source & Obs date & Exposure$^{a}$ & Count rate (0.3--10 keV) \\
       &          & (s)     & (counts s$^{-1}$) \\
\hline
\hline
V709 Cas          & Jul 15, 2007 & 290   & $0.245\pm0.029$    \\
                  & Jul 17, 2007 & 1156  & $0.415\pm0.019$    \\
                  & Nov 04, 2007 & 3970  & $0.497\pm0.011$    \\ 
                  & Nov 04, 2007 & 3532  & $0.396\pm0.011$    \\
                  & Nov 05, 2007 & 2370  & $0.453\pm0.014$    \\ 
                  & Nov 14, 2007 & 2985  & $0.451\pm0.012$    \\
                  & Apr 07, 2008 & 5732  & $0.437\pm0.009$    \\
                  & Apr 08, 2008 & 1376  & $0.437\pm0.020$    \\
                  & May 01, 2008 & 1651  & $0.430\pm0.016$     \\
                  & Jun 01, 2008 & 6365  & $0.492\pm0.009$     \\
total obs         &   --         & 29427 & $0.456\pm0.004$    \\
\hline
GK Per            & Dec 20, 2006 & 3931  & $0.883\pm0.015$    \\
                  & Dec 26, 2006 & 4493  & $0.445\pm0.010$    \\
                  & Jan 02, 2007 & 4679  & $0.482\pm0.010$    \\
                  & Jan 09, 2007 & 4802  & $0.455\pm0.010$    \\
                  & Jan 19, 2007 & 5993  & $0.460\pm0.009$     \\
                  & Jan 23, 2007 & 5756  & $0.480\pm0.009$     \\
                  & Jan 30, 2007 & 966   & $0.403\pm0.020$    \\
                  & Feb 04, 2007 & 3921  & $0.429\pm0.010$    \\
                  & Feb 08, 2007 & 5229  & $0.464\pm0.009$    \\
                  & Feb 12, 2007 & 2956  & $0.410\pm0.012$    \\
                  & Feb 16, 2007 & 6131  & $0.424\pm0.008$    \\
                  & Feb 19, 2007 & 2787  & $0.465\pm0.013$    \\
                  & Feb 23, 2007 & 3226  & $0.484\pm0.012$    \\
                  & Feb 26, 2007 & 6346  & $0.481\pm0.009$    \\
                  & Mar 02, 2007 & 3132  & $0.428\pm0.012$    \\
                  & Mar 06, 2007 & 7765  & $0.448\pm0.008$    \\
                  & Mar 09, 2007 & 4185  & $0.469\pm0.011$    \\
                  & Mar 13, 2007 & 5869  & $0.443\pm0.009$    \\
                  & Sep 27, 2007 & 2470  & $0.294\pm0.011$    \\
total obs         &      --      & 84637 & $0.475\pm0.002$    \\
\hline
IGR J06253+7334   & Nov 04, 2007 & 3411  & $0.114\pm0.006$    \\
                  & Dec 03, 2007 & 4468  & $0.114\pm0.005$    \\
                  & Dec 05, 2007 & 10092 & $0.123\pm0.003$    \\
                  & Dec 06, 2007 & 8928  & $0.084\pm0.003$    \\
total obs         &     --       & 26899 & $0.114\pm0.006$    \\
\hline
XSS J12270--4859  & Sep 15, 2005 & 4965  & $0.269\pm0.007$    \\
                  & Sep 24, 2005 & 1870  & $0.191\pm0.010$    \\
total obs         &   --         & 6835  & $0.251\pm0.006$    \\ 
\hline
IGR J14536--5522  & Oct 10, 2005 & 7165  & $0.395\pm0.008$    \\
                  & Mar 06, 2006 & 2040  & $0.321\pm0.013$    \\
total obs         &    --        & 9205  & $0.383\pm0.007$    \\
\hline
IGR J15479--4529  & Jun 22, 2007 & 359   & $0.405\pm0.034$    \\
                  & Jun 24, 2007 & 3978  & $0.369\pm0.010$    \\
                  & Jun 26, 2007 & 1002  & $0.289\pm0.017$    \\
                  & Jan 25, 2008 & 4772  & $0.485\pm0.010$    \\
total obs         &    --        & 10111 & $0.430\pm0.007$    \\
\hline
IGR J16167--4957  & Sep 09, 2005 & 2434  & $0.167\pm0.008$     \\
                  & Jan 26, 2006 & 3087  & $0.182\pm0.008$    \\
                  & Feb 15, 2007 & 5189  & $0.185\pm0.006$    \\
total obs         &   --         & 10710 & $0.184\pm0.004$    \\
\hline
IGR J16500--3307  & Jan 27, 2007 & 4594  & $0.136\pm0.005$    \\ 
\hline
\hline
\end{tabular}
\end{minipage}
\end{table*}
\begin{table*}
\begin{minipage}{155mm}
\centering
\begin{tabular}{lccc}
\hline
\hline
Source & Obs date & Exposure$^{a}$ & Count rate (0.3--10 keV) \\
       &          & (s)     & (counts s$^{-1}$) \\
\hline
\hline
IGR J17195--4100  & Oct 29, 2005 & 3474  & $0.434\pm0.011$    \\
                  & Oct 30, 2005 & 1588  & $0.455\pm0.017$    \\
                  & Jun 21, 2007 & 719   & $0.400\pm0.024$    \\
                  & Apr 30, 2008 & 7323  & $0.462\pm0.001$    \\
                  & May 02, 2008 & 2727  & $0.459\pm0.012$    \\
total obs         &    --        & 15831 & $0.456\pm0.005$    \\
\hline
IGR J17303--0601  & Feb 22, 2007 & 6520  & $0.369\pm0.008$    \\
                  & Feb 23, 2007 &11040  & $0.398\pm0.006$    \\
                  & Feb 25, 2007 & 4213  & $0.461\pm0.010$    \\
                  & Feb 28, 2007 & 817   & $0.548\pm0.026$    \\
                  & Mar 02, 2007 & 2043  & $0.456\pm0.015$    \\
total obs         &    --        &24633  & $0.411\pm0.004$    \\
\hline
FO Aqr            & May 04, 2006 & 9939  & $0.316\pm0.006$    \\
                  & May 04, 2006 & 3894  & $0.328\pm0.009$    \\
                  & May 04, 2006 & 1189  & $0.256\pm0.015$    \\
                  & May 04, 2006 & 5454  & $0.244\pm0.007$    \\
total obs         &   --         &20476  & $0.292\pm0.004$    \\
\hline
\hline 
\end{tabular}
\begin{list}{}{}
\item $^{a}$ Total on-source exposure time.
\end{list}
\end{minipage}
\end{table*}

\clearpage

\begin{landscape}
\begin{table}
\footnotesize
\caption{XRT/IBIS spectral analysis results of sources fitted with
the basic model (see text). Frozen parameters are written between squared brackets.}
\label{Tab3}
\begin{tabular}{lcccccccccc}
\hline
\hline
Source  & Energy band &  $N_{\rm H}$ & $N_{\rm H(pc)}$ & $C_{\rm F}$ & kT$_{\rm BB}$ &
kT$_{\rm Brems}$  & F$_{\rm 2-10~keV}$& F$_{\rm 20-100 ~keV}$& $C_{\rm calib}$$^{a}$ & $\chi^{2}/\nu$\\
  & (keV) &($10^{22}$ cm$^{-2}$)  & ($10^{22}$ cm$^{-2}$)  &  & (eV) & (keV)  & ($10^{-11}$ erg cm$^{-2}$ 
s$^{-1}$) & ($10^{-11}$ erg cm$^{-2}$ s$^{-1}$) &  &  \\
\hline
\hline
V709 Cas$^{b}$ &  0.3--100  & 0.23$^{+0.10}_{-0.04}$ & 12.0$^{+11.0}_{-5.6}$ & 0.28$^{+0.07}_{-0.06}$ & 
84$^{+6}_{-8}$ & 25.6$^{+2.7}_{-2.4}$ & 3.1 & 5.0 & $1.54^{+0.22}_{-0.22}$ & 325.8/329\\
\hline
GK Per$^{c}$ &  1--100  & 2.8$^{+0.24}_{-0.12}$ & 16.8$^{+2.4}_{-1.7}$ & 0.69$^{+0.03}_{-0.02}$ &
82$^{+6}_{-8}$ & 23.0$^{+9.2}_{-6.5}$ & 16.0 & 2.0 & $0.12^{+0.07}_{-0.04}$ & 510.5/499\\
\hline
IGR J06253+7334$^{d}$ & 0.3--100  & 0.06$^{+0.10}_{-0.01}$ & 6.7$^{+4.7}_{-3.0}$ & 
0.62$^{+0.08}_{-0.10}$ &
65$^{+8}_{-5}$ & $29.4^{+16.0}_{-9.0}$ & 0.7 & 1.1 & $1.16^{+0.92}_{-0.57}$ & 104.5/120 \\
\hline
XSS J12270--4859&  0.4--100  & 0.037$^{+0.025}_{-0.023}$ & -- &  -- &  $<$ 127 & 33.4$^{+18.4}_{-10.6}$ & 
1.2 & 2.3 & $1.60^{+0.88}_{-0.56}$ & 75.3/76 \\
\hline
IGR J14536--5522 & 0.3--100 & 0.071$^{+0.07}_{-0.03}$ & 12.2$^{+8.3}_{-6.0}$ & 0.58$^{+0.08}_{-0.09}$ 
& 60$^{+7}_{-5}$ & 16.3$^{+5.0}_{-4.0}$ & 2.4 & 1.5 & $1.03^{+0.65}_{-0.40}$ & 130.4/140 
\\
\hline
IGR J15479--4529 & 0.3--100 & 0.065$^{+0.014}_{-0.008}$ & 14.6$^{+8.3}_{-3.7}$ & 0.52$^{+0.08}_{-0.07}$ 
& 101$^{+12}_{-10}$ & 28.4$^{+3.2}_{-2.8}$ & 2.4 & 6.1 & $1.82^{+0.36}_{-0.34}$ & 116.4/108 \\
\hline
IGR J16167--4957 & 0.6--100 & 0.96$^{+0.15}_{-0.12}$ & 94.0$^{+42.7}_{-37.2}$ & 0.84$^{+0.11}_{-0.34}$ 
& 95$^{+15}_{-12}$ & 16.7$^{+6.6}_{-4.5}$ & 1.7 & 2.3 & $0.75^{+1.74}_{-0.36}$ & 94.0/86 \\
\hline
IGR J16500--3307 & 0.9--50 & 1.88$^{+0.73}_{-0.64}$ & -- & -- &
119$^{+28}_{-22}$ & 32.0$^{+31}_{-11}$ & 1.1 & 1.9 & $1.25^{+0.81}_{-0.52}$ & 46.7/30 \\
\hline
IGR J17195--4100 & 0.3--100 & 0.23$^{+0.04}_{-0.06}$ & --  & --  &
104$^{+12}_{-11}$ & 29.7$^{+5.4}_{-4.2}$ & 2.4 & 3.6 & $1.35^{+0.31}_{-0.26}$ & 263.3/263 \\
\hline
IGR J17303--0601$^{e}$ & 0.3--100 & 0.20$^{+0.04}_{-0.02}$ & 45.2$^{+25.1}_{-18.6}$ & 
0.52$^{+0.14}_{-0.12}$ & 85$^{+9}_{-12}$ & 31.6$^{+12.7}_{-7.8}$ & 1.9 & 4.4 & 
$1.22^{+0.59}_{-0.49}$ & 320.0/290 \\
\hline
FO Aqr           & 0.4--50  & 0.78$^{+0.19}_{-0.18}$ & 7.2$^{+1.2}_{-0.9}$ & 0.82$^{+0.04}_{-0.03}$
& 61$^{+8}_{-6}$ & [29.7]$^{f}$ &  3.6  &  2.1  & $0.75^{+0.22}_{-0.21}$ & 155.7/169 \\  
\hline
\hline
\end{tabular}
$^{a}$ XRT/IBIS cross-calibration constant;\\
$^{b}$ For this source the fit also includes a narrow Gaussian line at $E=6.46^{+0.05}_{-0.04}$ keV
with $EW = 141^{+43}_{-58}$ eV (see text for details);\\ 
$^{c}$ For this source the fit also includes a Gaussian iron line at $E=6.35^{+0.05}_{-0.03}$ keV 
with $\sigma = 0.29^{+0.05}_{-0.04}$ keV and $EW = 371^{+59}_{-53}$ eV (see text for details);\\
$^{d}$ For this source the fit also includes a narrow Gaussian line at 6.4 keV with 
$EW = 195^{+115}_{-108}$ eV (see text for details);\\
$^{e}$ For this source the fit also includes a further thermal component (MEKAL code)
with $kT = 0.21^{+0.01}_{-0.02}$ keV (see text for details);\\
$^{f}$ Fixed at the value obtained by fitting the IBIS spectrum alone.
\end{table}
\end{landscape}

\clearpage

\begin{table*}
\begin{center}
\footnotesize
\caption{Fit statistics for each source with and without the blackbody component.}
\label{Tab4}
\begin{tabular}{lccc}
\hline
\hline
Source  & \multicolumn{2}{c}{$\chi^{2}/\nu$}& $F$--test \\
        & No BB    &  with BB  &     \\
\hline
\hline
V709 Cas & 349.2/331 & 325.8/329 & $1.12\times 10^{-5}$ \\
\hline
GK Per  &  615.2/501 & 510.5/499 & $6.10\times 10^{-21}$ \\
\hline
IGR J06253+7334 & 173.4/122 & 104.5/120 & $6.37\times 10^{-14}$ \\
\hline
XSS J12270--4859 &  81.2/78 & 75.3/76 & $5.69\times 10^{-2}$ \\  
\hline
IGR J14536--5522 &  434.4/142  &  130.4/140 & $2.66\times 10^{-37}$ \\ 
\hline
IGR J15479--4529 &  182.1/108 & 116.1/106 & $4.36\times10^{-11}$ \\
\hline
IGR J16167--4957 &  103.0/88  & 94.0/86 & $1.96\times10^{-2}$ \\
\hline
IGR J16500--3307 &  64.2/32 & 46.7/30 & $8.45\times10^{-3}$ \\
\hline
IGR J17195--4100 &  277.0/265 & 263.3/263 & $1.27\times10^{-3}$ \\
\hline
IGR J17303--0601 &  417.4/292 & 320.0/290 & $1.85\times10^{-17}$ \\
\hline
FO Aqr &  190.3/171 & 155.7/169 & $4.32\times10^{-8}$ \\
\hline
\hline
\end{tabular}
\end{center}
\end{table*}

\begin{table*}
\begin{center}
\footnotesize
\caption{Table of characteristics of the CVs observed by \emph{INTEGRAL}.}
\label{Tab5}
\begin{tabular}{lcccccc}
\hline
\hline
Source  & P$_{\rm orb}$ & P$_{\rm spin}$ & P$_{\rm spin}$/P$_{\rm orb}$ & Mass & Dist  &  Refs \\
        & (min)     & (s)    &      &   (M$_{\sun}$)  &  (pc)   &    \\
\hline
\hline
IGR J00234+6141  &  $242.0\pm0.3$ & $563.53\pm0.62$  &  0.04  &  -- &  530     & 1  \\ 
\hline 
V709 Cas & 320.4 & 312.7 & 0.0167  & $0.90\pm0.10$  &  $230\pm20$  &  2, 3 \\   
\hline
GK Per  &  2875.2 & 351.34 &  0.00204 & $0.92^{+0.39}_{-0.13}$ & 420   & 2, 4  \\
\hline
BY Cam & 201.3  &  11846.4 &  0.981  &  $\geq 0.8$  &  190  &  2, 5  \\
\hline
IGR J06253+7334 & 283 & 1186.7 &  0.07  &  0.8 & 500  &  2, 6, 7  \\
\hline
XSS J12270--4859&  -- &  --  &  --  &   --  &  220 & 8  \\
\hline
V834 Cen &  101.4  &  --  &  --  & $0.70^{+0.20}_{-0.15}$  &  80  &  2, 9 \\
\hline
IGR J14536--5522 &  --  &  --  &  --  &  --  &  190  &  8  \\
\hline
IGR J15094-6649 &  --  &  --  &  --  &  --  &  140  &  8  \\
\hline 
IGR J15479--4529 &  562  &  693  &  0.021  &  $\geq 0.5$  & 540--840  &  2, 9  \\
\hline
IGR J16167--4957 & --  &  --  &  --  &  --  &  170  &  8  \\
\hline
IGR J16500--3307 & -- & -- & -- & -- &  $\sim$210  &  10 \\
\hline
V2400 Oph  &   205.2  &  927  &  0.075  & $0.69^{+0.06}_{-0.24}$ & 300  &  2, 4  \\
\hline
IGR J17195--4100 & --  &  --  &  --  &  --  &  110  &  8  \\
\hline
IGR J17303--0601 &  924  &  128  &  0.0023  &  0.89--1.02 &  --   & 11  \\  
\hline
V2487 Oph & --  &  --  &  --  &  --  &  --  &  12    \\
\hline
V1223 Sgr &  201.9 & 745.6 & 0.062 & $1.046^{+0.049}_{-0.012}$  &  527 & 2, 13   \\
\hline
V1432 Aql & 201.94  &  12150.4 &  1.003 & $1.2\pm0.2$  &  230  &  2, 14    \\
\hline
V2069 Cyg & 448.8  &  --  &  -- &  --  & 1650  &  2 \\
\hline
IGR J21335+5105 & 431.6 &  570.8   &  0.022 & 0.6--1.0 &  1400  &  2, 15 \\
\hline
SS Cyg  & 396.2 &  --  &  --  &  -- & 166 & 2  \\
\hline
FO Aqr & 291 & 1254.5 & 0.0718 & $1.19^{+0.11}_{-0.31}$ & 400 &  3, 16 \\
\hline
\hline
\end{tabular}
\end{center}
\begin{list}{}{}
\item References: [1] Bonnet-Bidaud et al. (2007); [2] Barlow et al. (2006) and references 
therein; [3] Suleimanov et al. (2005);
[4] Evans \& Hellier (2007); [5] Schwarz et al. (2005); [6] Staude et al. (2003); [7] Araujo-Betancor et 
al. (2003); 
[8] Masetti et al. (2006); [9] De Martino et al. (2006b); 
[10] Masetti et al. (2008); [11] De Martino et al. (2008); [12] Hernanz \& Sala (2002); [13] Beuermann et 
al. (2004); [14] Rana et al. (2005); [15] Bonnet-Bidaud et al. (2006); [16] Norton et al. (2004).
\end{list}
\end{table*}

\clearpage

\clearpage

\begin{figure*}
\centering
\includegraphics[width=0.35\linewidth,angle=-90]{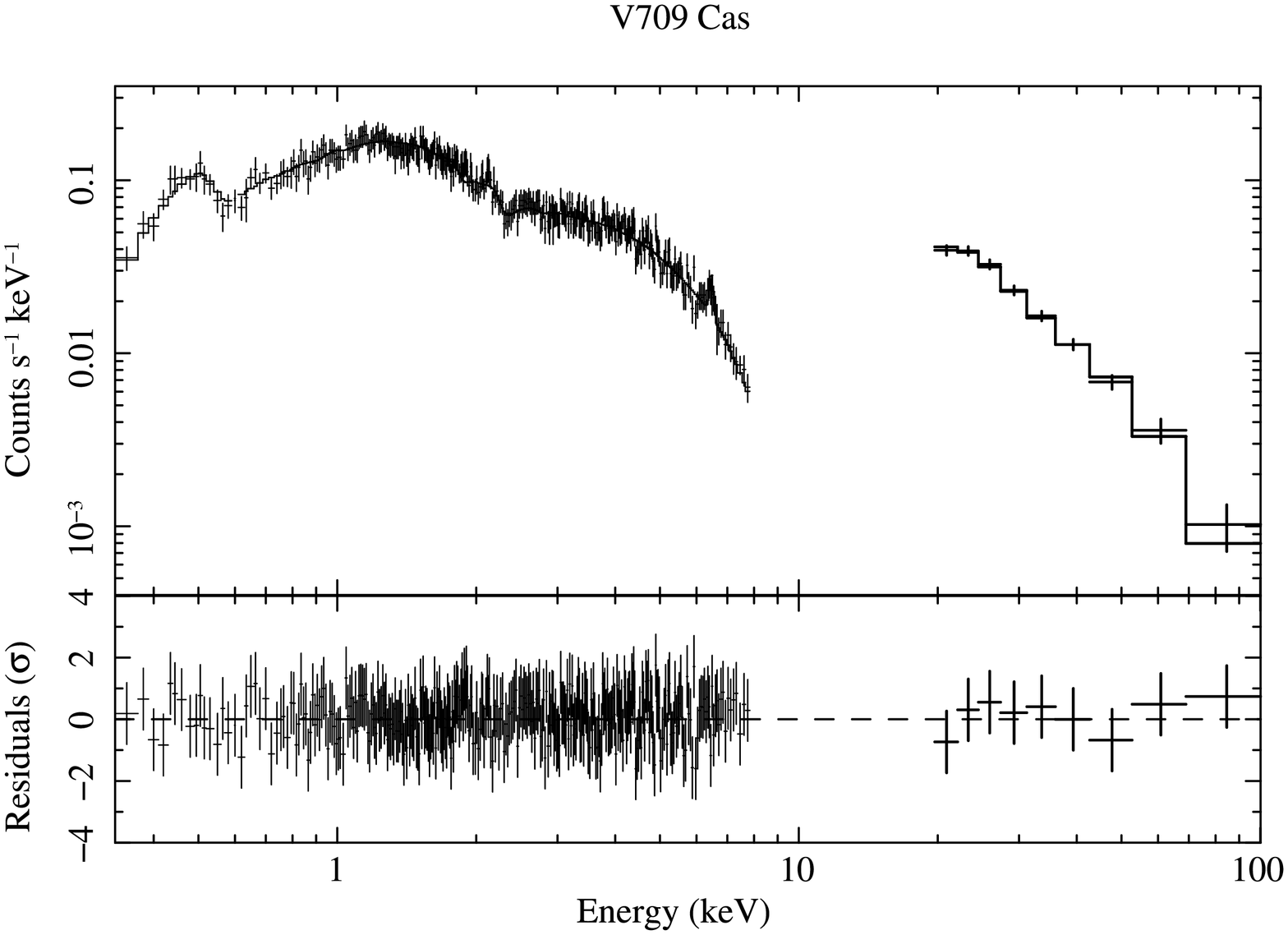}%
\includegraphics[width=0.35\linewidth,angle=-90]{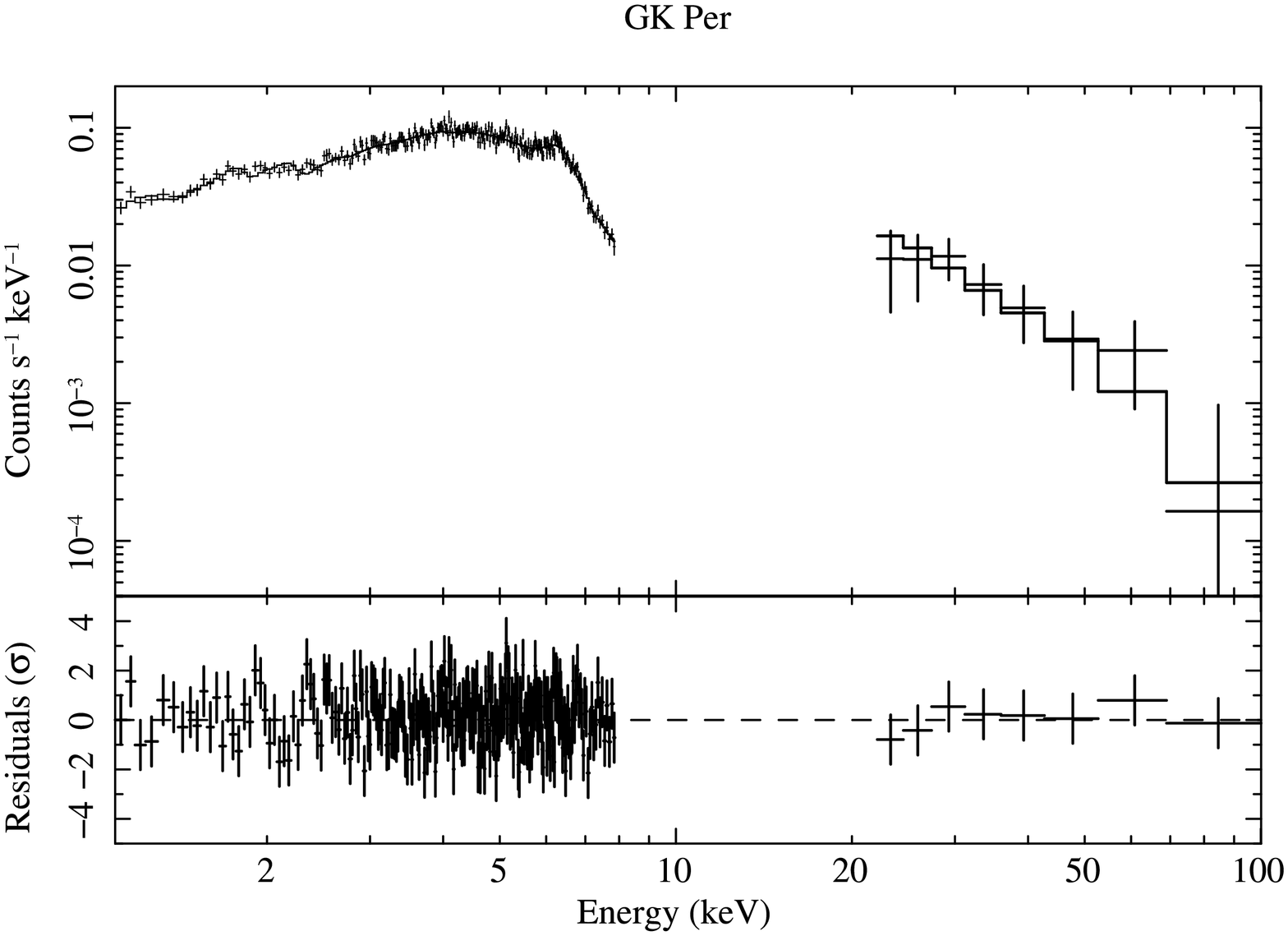}%
\vspace{8mm}

\includegraphics[width=0.35\linewidth,angle=-90]{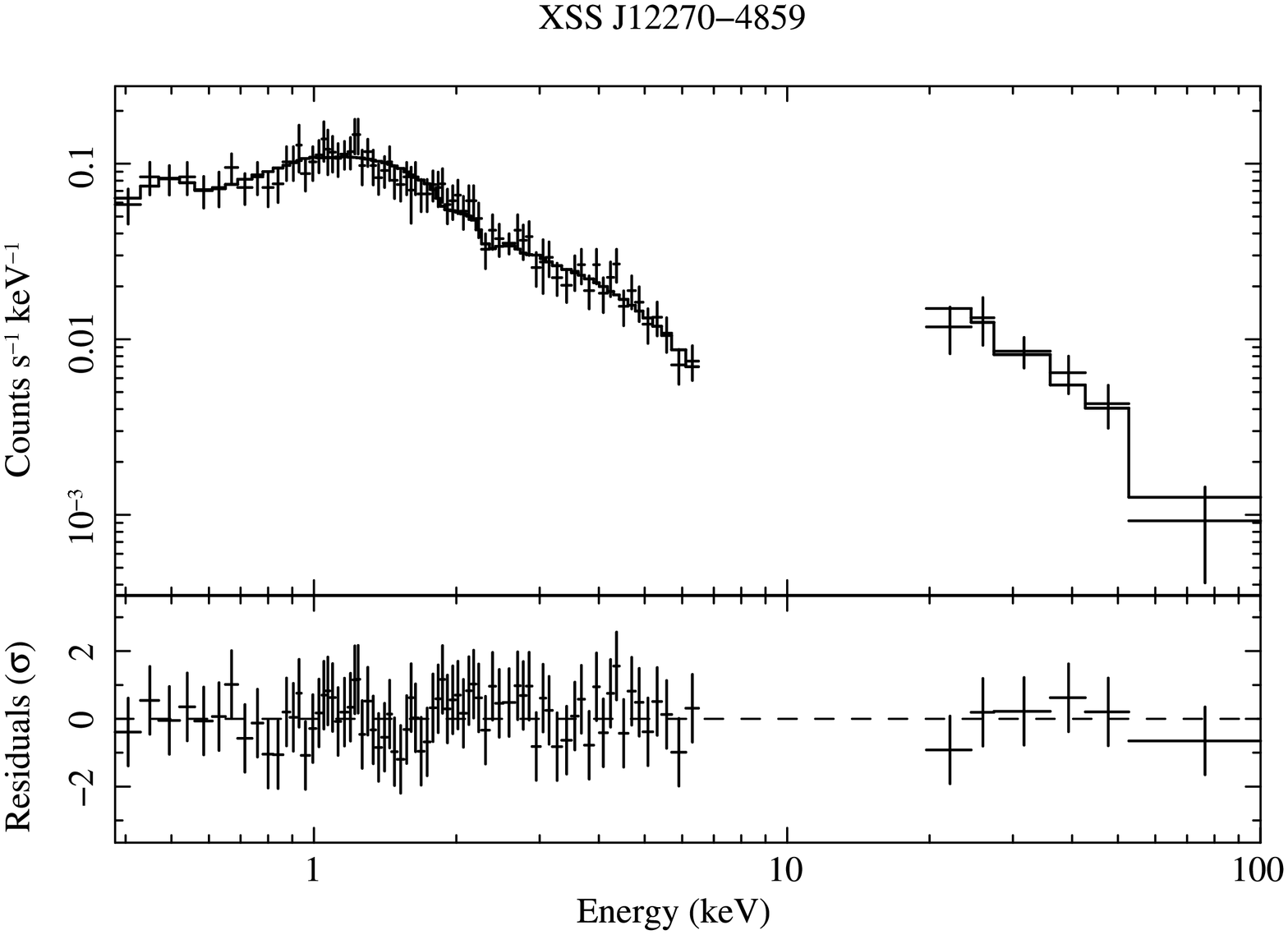}%
\includegraphics[width=0.35\linewidth,angle=-90]{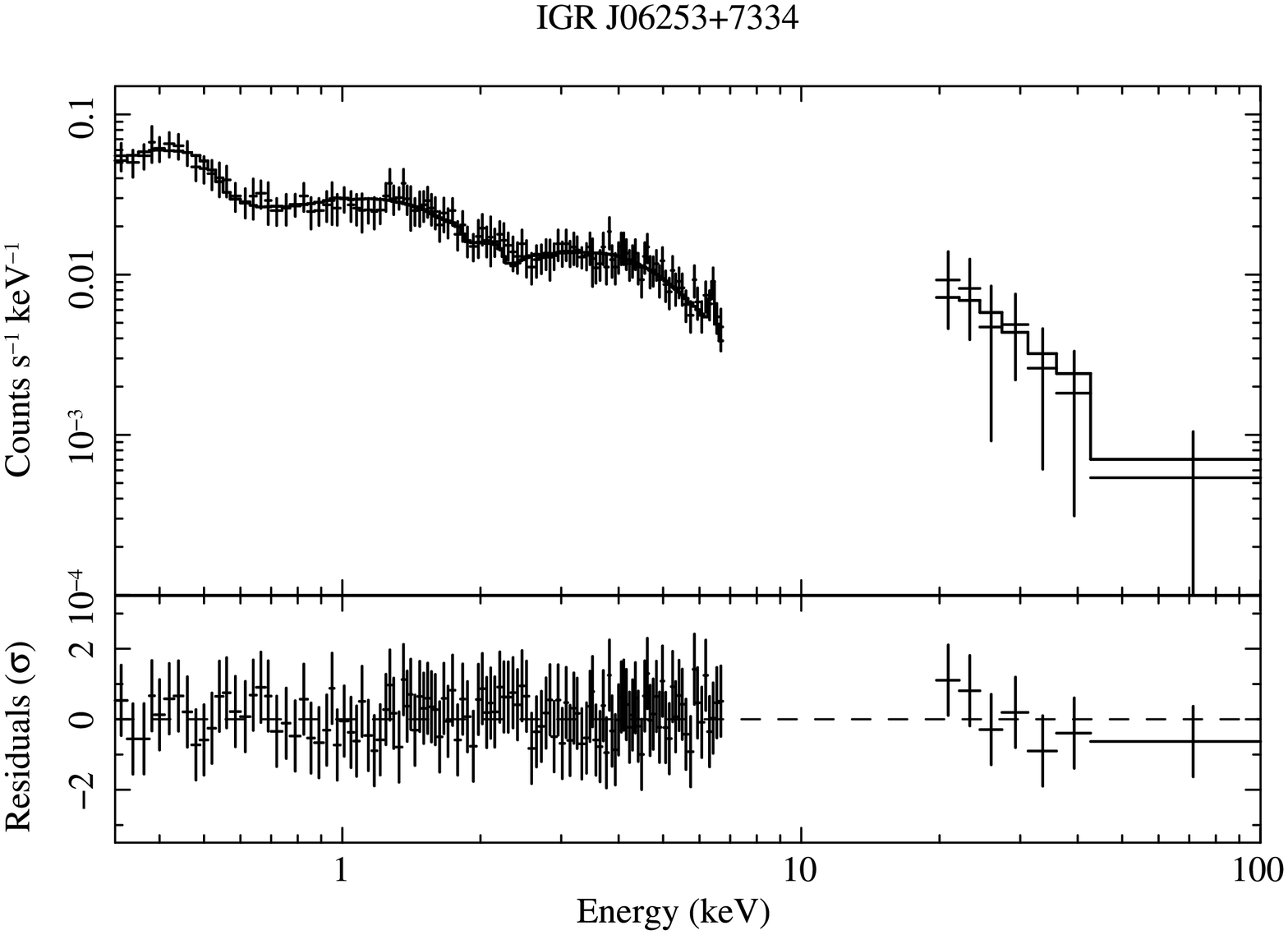}%
\vspace{8mm}

\includegraphics[width=0.35\linewidth,angle=-90]{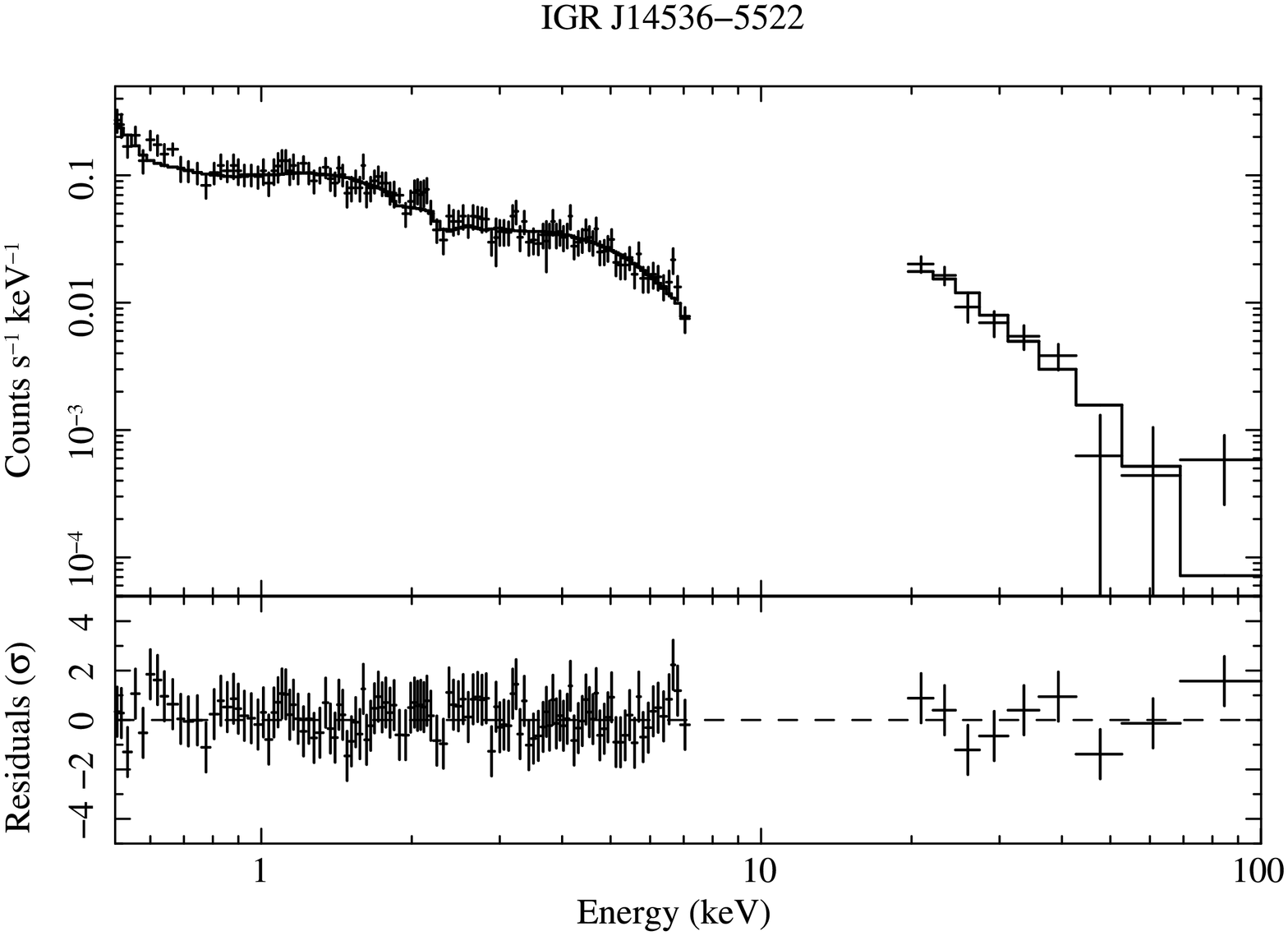}%
\includegraphics[width=0.35\linewidth,angle=-90]{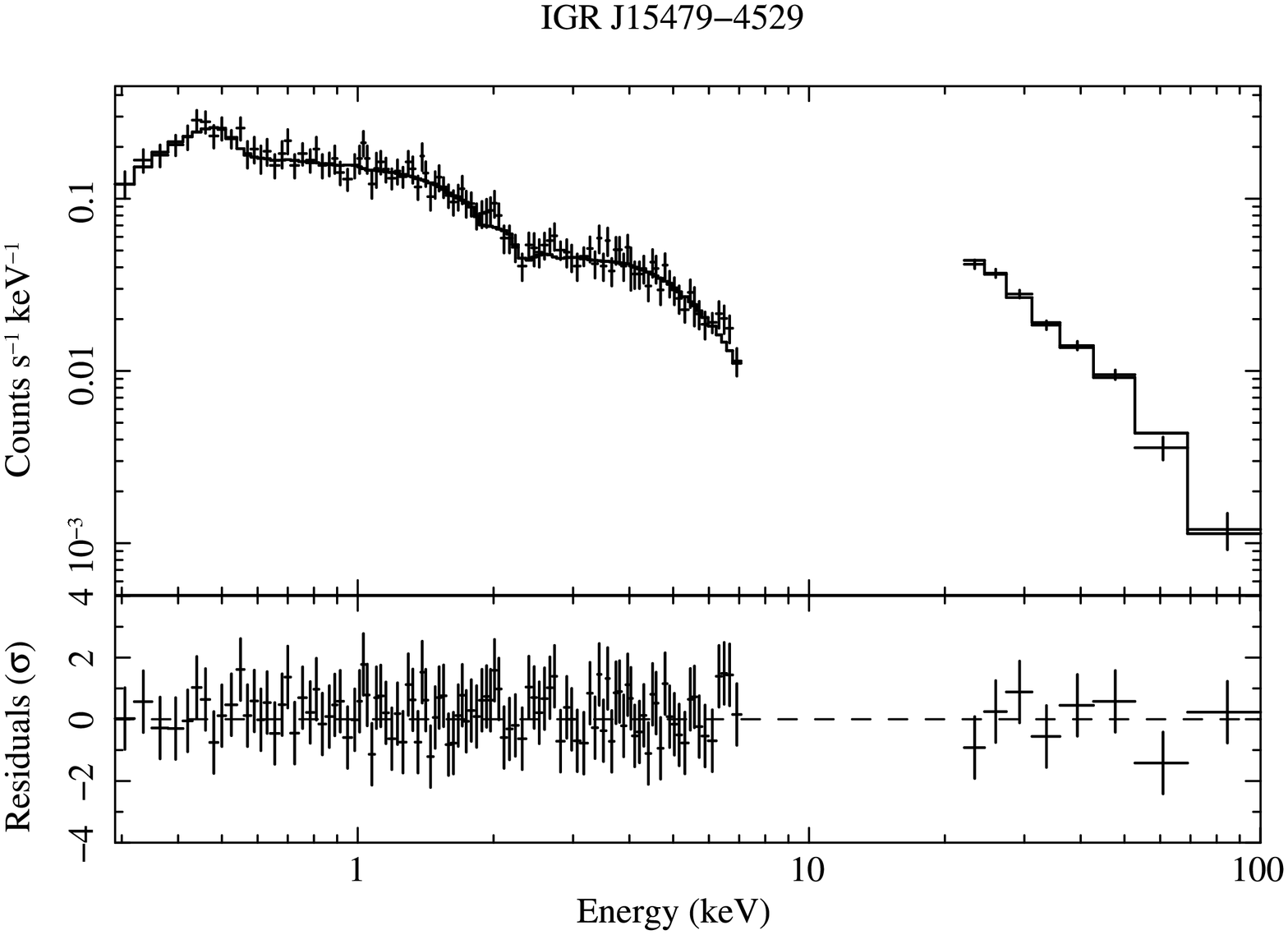}%
\caption{Best-fit model of the XRT/IBIS phase-averaged spectrum of the eleven systems followed-up by 
\emph{Swift}, as described in Table~\ref{Tab3} (upper 
panel); residuals to this model are in units of $\sigma$ (lower panel).}
\label{fig5}
\end{figure*}

\clearpage

\setcounter{figure}{4}

\begin{figure*}
\centering
\includegraphics[width=0.35\linewidth,angle=-90]{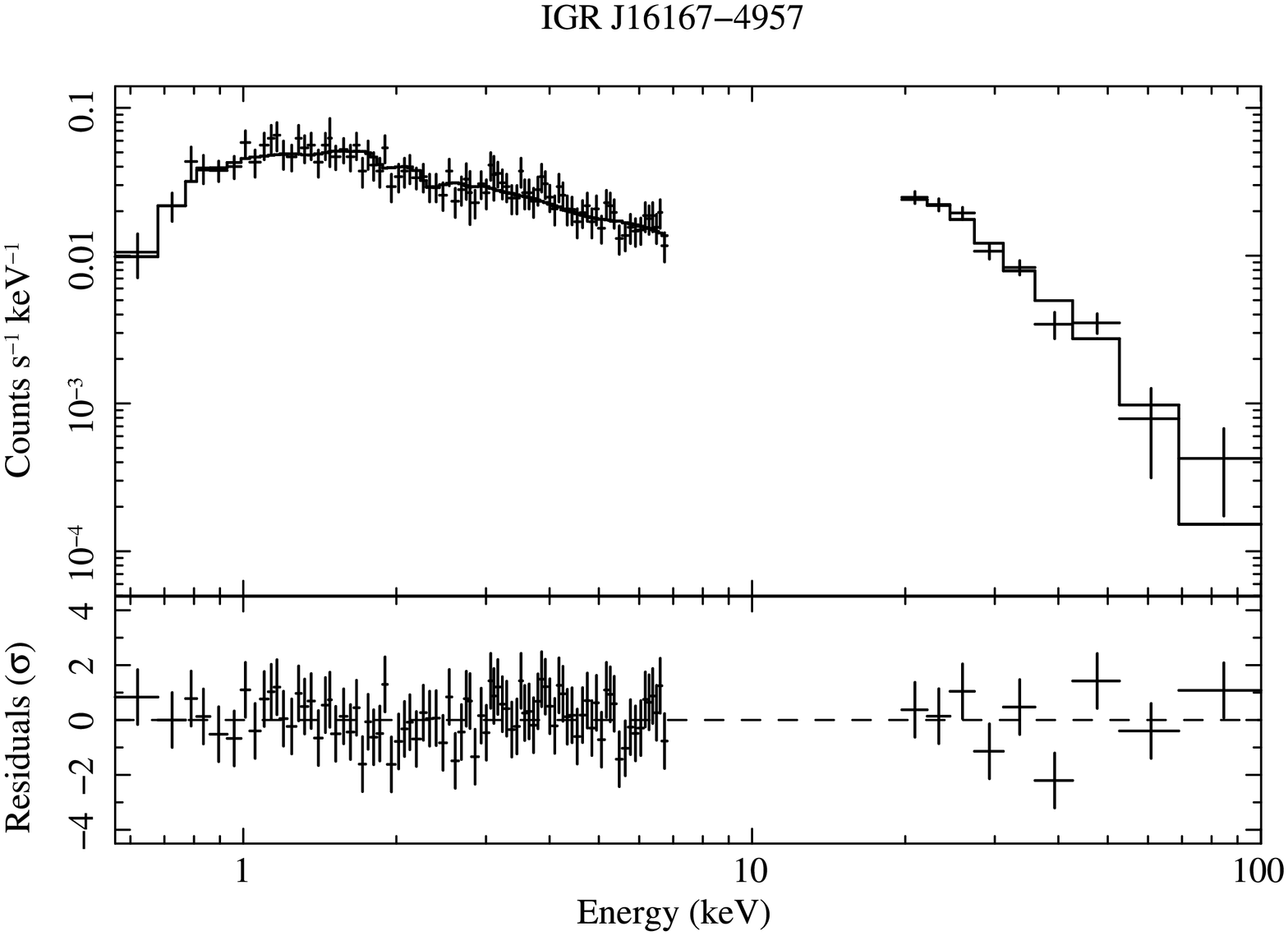}%
\includegraphics[width=0.35\linewidth,angle=-90]{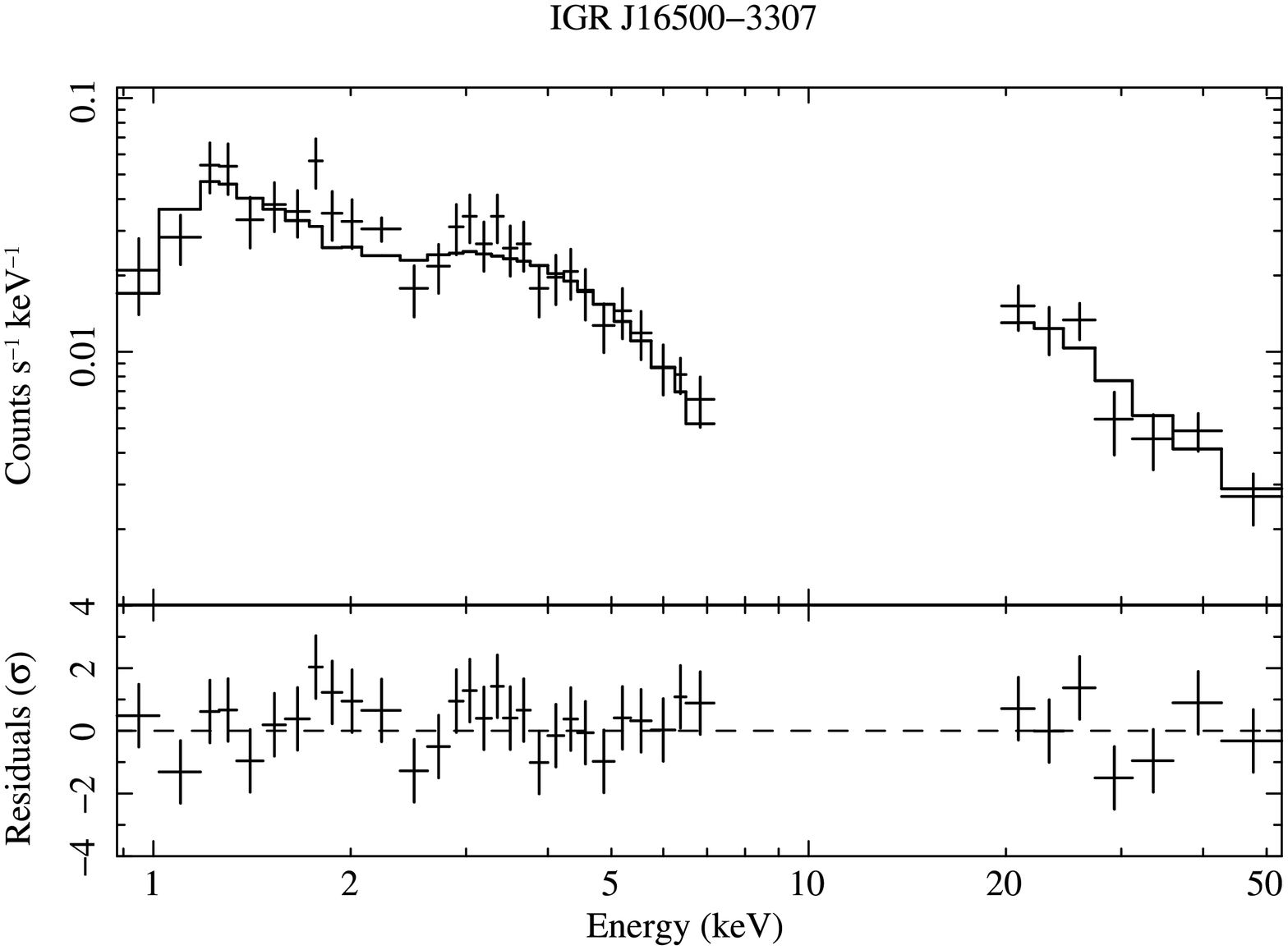}%
\vspace{8mm}

\includegraphics[width=0.35\linewidth,angle=-90]{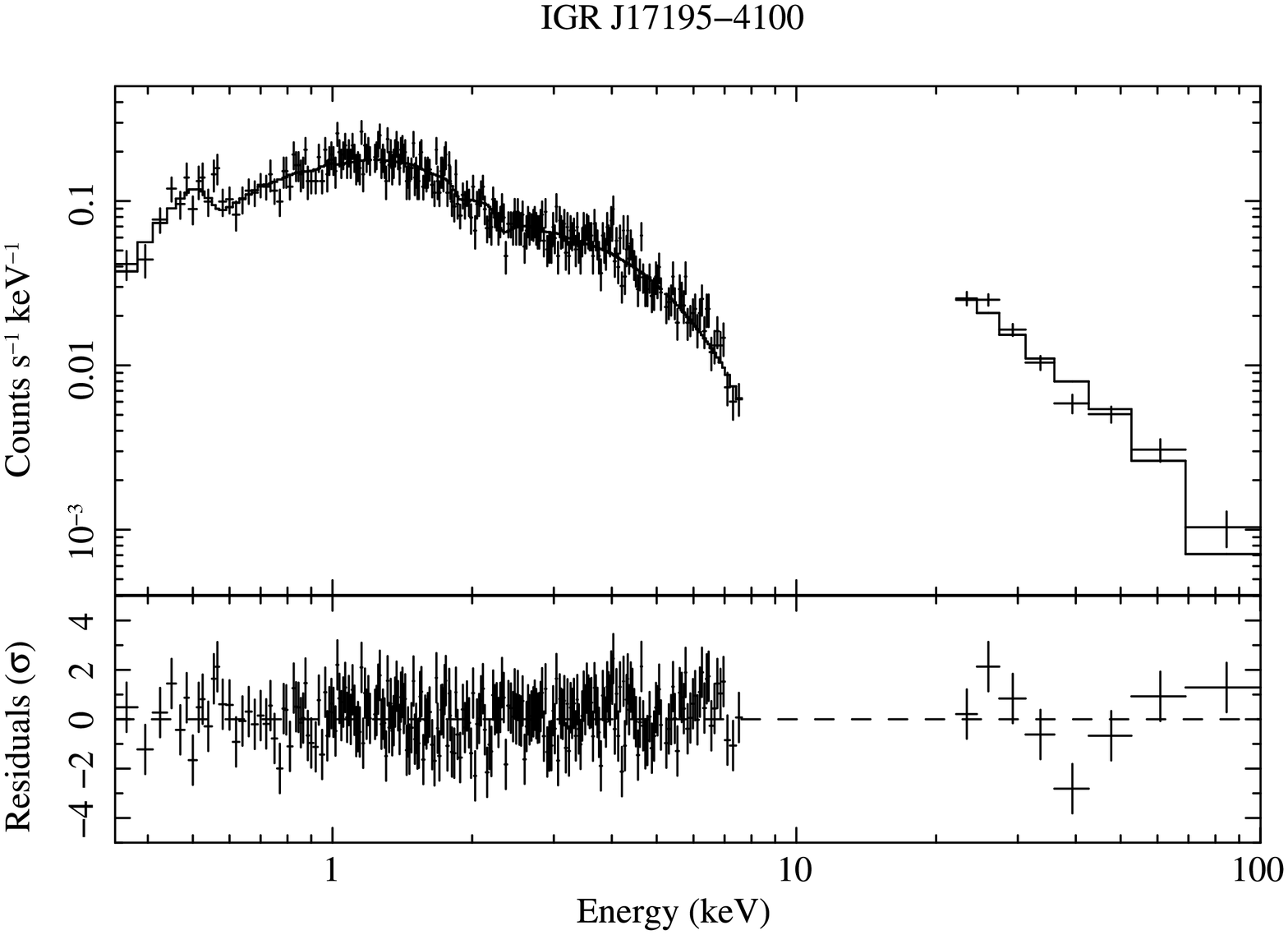}%
\includegraphics[width=0.35\linewidth,angle=-90]{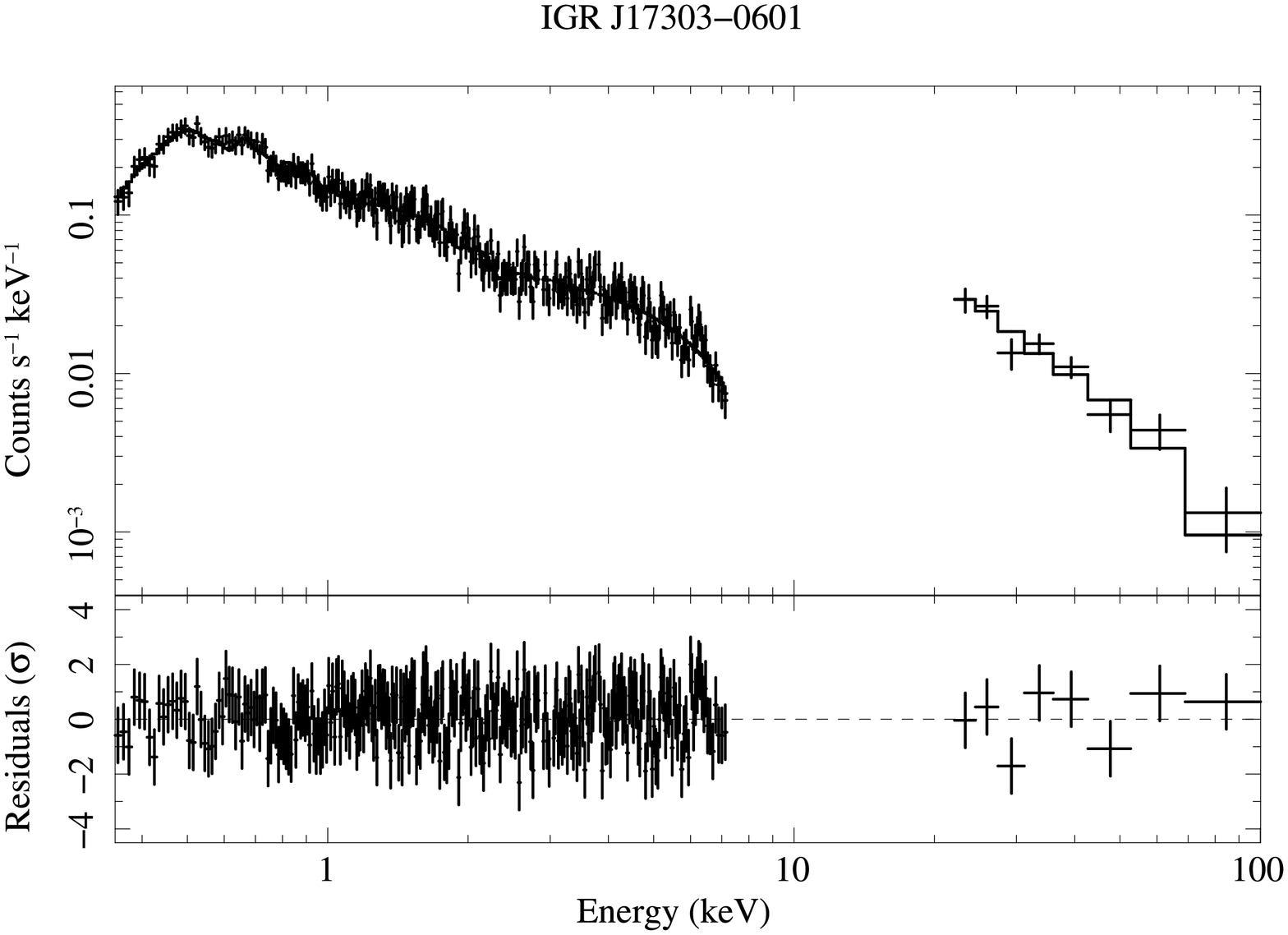}%
\vspace{8mm}

\includegraphics[width=0.35\linewidth,angle=-90]{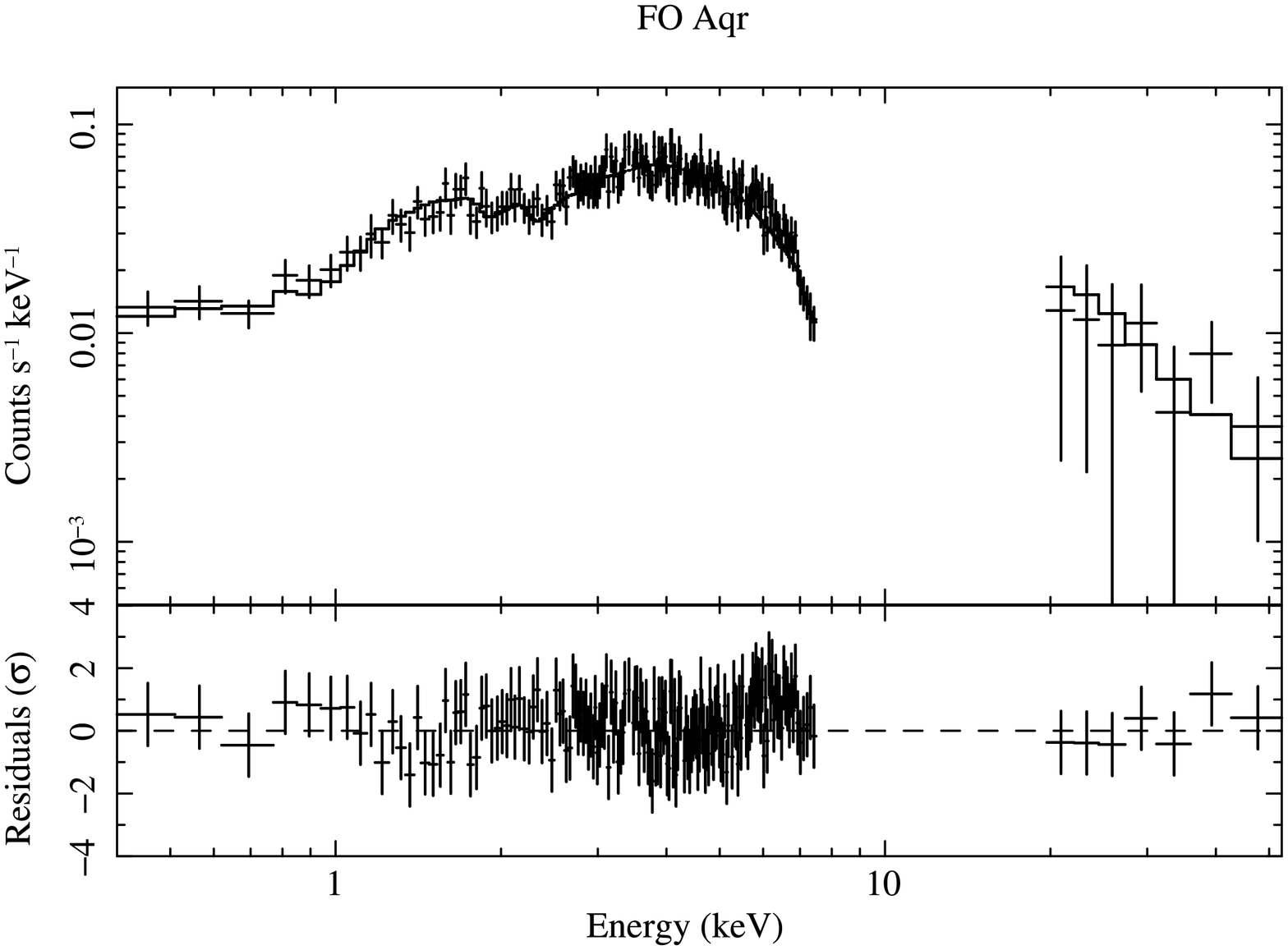}%
\caption{\emph{-Continued}}
\end{figure*}

\clearpage

\begin{figure*}
\includegraphics[width=0.35\linewidth,angle=-90]{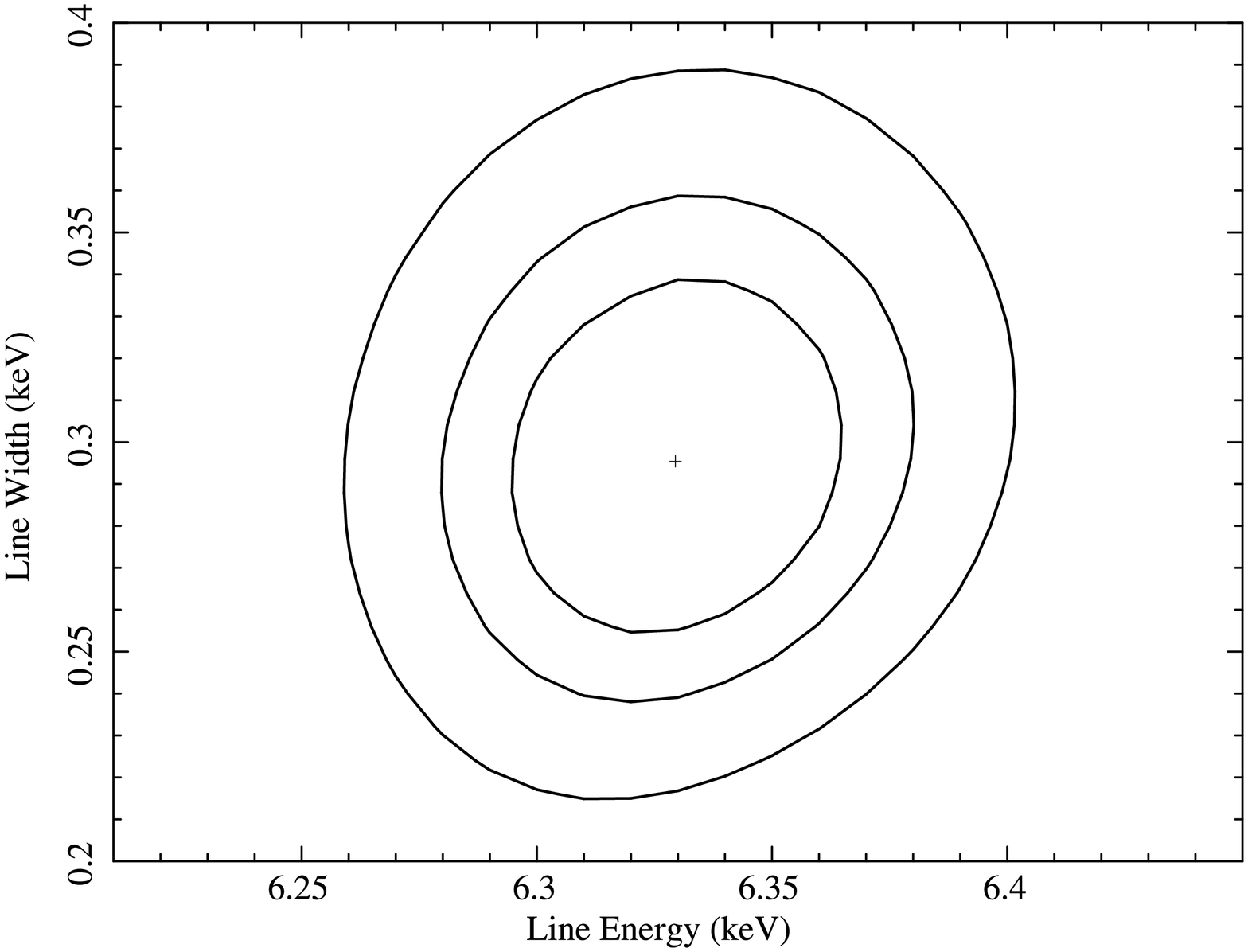}
\caption{Confidence contours (at 68\%, 90\%, and 99\% confidence level) 
of the iron line energy versus line width for the best-fit model
of GK Per.}
\label{fig6}
\end{figure*}


\begin{thebibliography}{99}
\bibitem[\protect\citeauthoryear{Araujo-Betancor et al.}{2003}]{b1} Araujo-Betancor, S., 
G\"{a}nsicke, B. T, Hagen, H.--J, Rodriguez-Gil, P., \& Engels, D. 2003, A\&A, 406, 213
\bibitem[\protect\citeauthoryear{Arnaud}{1996}]{b2}
Arnaud K. A. 1996, XSPEC: the first ten years, in Astronomical Data Analysis Software and
Systems V, ed. G. H. Jacoby, \& J. Barnes, ASP Conf Ser. 101, 17
\bibitem[\protect\citeauthoryear{Barlow et al.}{2006}]{b3} Barlow, E. J., Knigge, C., Bird A. J., et al.
2006, MNRAS, 372, 224
\bibitem[\protect\citeauthoryear{Beuermann et al.}{2004}]{b4} 
Beuermann, K., Harrison, Th. E., McArthur, B. E., Benedict, G. F., \& G\"{a}nsicke, B. T.
2004, A\&A, 419, 291
\bibitem[\protect\citeauthoryear{Bird et al.}{2007}]{b5} Bird, A. J., Malizia, A., Bazzano, A., et 
al. 2007, ApJS, 170, 175 
\bibitem[\protect\citeauthoryear{Bikmaev et al.}{2006}]{b6} Bikmaev, I. F., Revnivtsev, M. G., Burenin, 
R. A. \& Sunyaev, R. A. 2006, Astr. Letters, 32, 588
\bibitem[\protect\citeauthoryear{Bonnet-Bidaud et al.}{2007}]{b7} Bonnet-Bidaud, J.--M., De Martino, D., 
Falanga, M., Mouchet, M., \& Masetti N. 2007, A\&A, 473, 185
\bibitem[\protect\citeauthoryear{Bonnet-Bidaud et al.}{2006}]{b8} Bonnet-Bidaud, J.--M., Mouchet, M.,
De Martino, D., Silvotti, R., \& Motch, C 2006, A\&A, 445, 103
\bibitem[\protect\citeauthoryear{Burrows et al.}{2005}]{b9} Burrows, D. N., Hill, J. E., Nousek, J. A., 
et al. 2005, Space Sci. Rev., 120, 165
\bibitem[\protect\citeauthoryear{Cropper}{1990}]{b10} Cropper, M. 1990, Space Sci. Rev., 54, 195
\bibitem[\protect\citeauthoryear{De Martino et al.}{2008}]{b11} De Martino, D., Matt, G. Mukai, K., et 
al. 2008, A\&A, 481, 149
\bibitem[\protect\citeauthoryear{De Martino et al.}{2006a}]{b12} De Martino, D., Matt, G., Mukai, K.,
Bonnet-Bidaud, J.--M., Burwitz, V., G\"{a}nsicke, B. T., Haberl, F., \& Mouchet, M. 2006a, A\&A, 454, 
287
\bibitem[\protect\citeauthoryear{De Martino et al.}{2006b}]{b13} De Martino, D., Bonnet-Bidaud, J.--M., 
Mouchet, M., G\"{a}nsicke, B. T., Haberl, F., \& Motch, C. 2006b, A\&A, 449, 115
\bibitem[\protect\citeauthoryear{De Martino et al.}{2004a}]{b14} De Martino, D., Matt, G., Belloni, T., 
Chiappetti, L., Haberl, F., \&  Mukai, K. 2004a, Nuc. Phys. B Proc. Supp., 132, 693
\bibitem[\protect\citeauthoryear{De Martino et al.}{2004b}]{b15} De Martino, D., Matt, G., Belloni, T., 
Haberl, F. \& Mukai, K. 2004b, A\&A, 415, 1009
\bibitem[\protect\citeauthoryear{De Martino et al.}{2001}]{b16} De Martino, D., Matt, G. Mukai, K., et 
al. 2001, A\&A, 377, 499
\bibitem[\protect\citeauthoryear{Evans \& Hellier}{2007}]{b17} Evans, P. E. \& Hellier, C. 2007, 
ApJ, 663, 1277
\bibitem[\protect\citeauthoryear{Falanga et al.}{2005}]{b18} Falanga, M., Bonnet-Bidaud, J. M. \& 
Suleimanov, V. 2005, A\&A, 444, 561
\bibitem[\protect\citeauthoryear{Gehrels et al.}{2004}]{b19} Gehrels, N., Chincarini, G., Giommi, 
P. et al. 2004, ApJ, 611, 1005
\bibitem[\protect\citeauthoryear{G\"{a}nsicke et al.}{2005}]{b20} G\"{a}nsicke, B. T., Marsh, 
T. R., Edge, A., et al. 2005, Atel 463
\bibitem[\protect\citeauthoryear{Goldwurm et al.}{2003}]{b21} Goldwurm, A., David, P., 
Foschini, L., et al. 2003, A\&A, 411, L223
\bibitem[\protect\citeauthoryear{Hellier et al.}{2004}]{b22} Hellier, C., \& Mukai, K. 2004, MNRAS, 352, 
1037
\bibitem[\protect\citeauthoryear{Hellier}{1991}]{b23} Hellier, C. 2001, Cataclysmic Variables Stars 
(Springer-Praxis: Chichester)
\bibitem[\protect\citeauthoryear{Hellier}{1991}]{b24} Hellier, C. 1991, MNRAS, 251, 693
\bibitem[\protect\citeauthoryear{Hernanz \& Sala}{2002}]{b25} Hernanz, M., \& Sala, G. 2002, Science, 
298, 393
\bibitem[\protect\citeauthoryear{Hill et al.}{2004}]{b26} Hill, J. E., Burrows, D. N., Nousek, J. A., 
et al. 2004, Proc. SPIE, 5165, 217
\bibitem[\protect\citeauthoryear{Ishida et al.}{1994}]{b27} Ishida, M., Mukai, K, \& Osborne J.P.
1994, PASJ, 46, L81
\bibitem[\protect\citeauthoryear{Konig et al.}{2006}]{b28} K\"{o}nig, M., Beuermann, K., \& 
G\"{a}nsicke, B. T. 2006, A\&A, 449, 1129
\bibitem[\protect\citeauthoryear{Masetti et al.}{2008}]{b30} Masetti, N., Mason, E., Morelli, L., et al. 
2008, A\&A, 482, 113
\bibitem[\protect\citeauthoryear{Masetti et al.}{2006}]{b31} Masetti, N., Morelli, L., Palazzi, E., 
et al. 2006, A\&A 459, 21
\bibitem[\protect\citeauthoryear{Moretti et al.}{2004}]{b32} Moretti, A., Campana, S., Tagliaferri, G., 
et al. 2004, Proc. SPIE, 5165, 232 
\bibitem[\protect\citeauthoryear{Norton et al.}{2004}]{b33} Norton, A. J., Wynn, G. A., \& Somerscales, 
R. V. 2004, ApJ, 614, 349
\bibitem[\protect\citeauthoryear{Pringle \& Savonije}{1979}]{b34} Pringle, J. E. \& Savonije, G. J. 1979, 
MNRAS, 187, 777
\bibitem[\protect\citeauthoryear{Ramsay et al.}{2008}]{b35} Ramsay, G., Wheatley, P. J., 
Norton, A. J., Hakala, P, \& Baskill, D. 2008, MNRAS, 387, 1162
\bibitem[\protect\citeauthoryear{Rana et al.}{2005}]{b36} Rana, V. R., Singh, K. P., Barrett, P. E., \&
Buckley, D. A. H. 2005, ApJ, 625, 351
\bibitem[\protect\citeauthoryear{Ritter \& Kolb}{2003}]{b37} Ritter H. \& Kolb U. 2003, A\&A, 404, 301 
(update RKcat7.10) 
\bibitem[\protect\citeauthoryear{Romano et al.}{2006}]{b38} Romano, P., Campana, S., Chincarini, G., et 
al., 2006, A\&A, 456, 917
\bibitem[\protect\citeauthoryear{Staude et al.}{2003}]{b39} Staude, A., Schwope, A. D., Krumpe, M.,
Hambaryan, V., \& Schwarz, R. 2003, A\&A, 406, 253 
\bibitem[\protect\citeauthoryear{Suleimanov et al.}{2005}]{b40} Suleimanov, V., Revnivtsev, M., \& 
Ritter, H. 2005, A\&A, 435, 191
\bibitem[\protect\citeauthoryear{Schwarz et al.}{2005}]{b41} Schwarz, R., Schwope, A. D., Staude, A., 
\& Remillard, R. A. 2005, A\&A, 444, 213
\bibitem[\protect\citeauthoryear{Schwarz et al.}{2005}]{b42} Ubertini, P., Lebrun, F., Di Cocco, G., et 
al. 2003, A\&A, 441, L131
\bibitem[\protect\citeauthoryear{Vrielmann}{2005}]{b43} Vrielmann, S., Ness, J.--U, \& Schmitt, H. H. 
M. 2005, A\&A, 439, 287 
\bibitem[\protect\citeauthoryear{Warner}{1995}]{b44} Warner, B. 1995, Cataclysmic Variables, 
(Cambridge: Cambridge Univ. Press)
\end{thebibliography}
\end{document}